\documentclass[useAMS,usenatbib]{mn2e}
\usepackage{graphicx}
\usepackage{rotating}
\usepackage{longtable}
\usepackage{lscape}
\usepackage{amssymb,amsmath}
\usepackage{url}
\usepackage{hyperref}
\usepackage{natbib}
\usepackage{fixltx2e}
\begin{document}

\title[Low z QSO properties]
{Low redshift quasars in the SDSS Stripe 82. The local environments.\\}
\author[Karhunen et al.]{
	K. ~Karhunen$^{1}$,
	J.~K.~Kotilainen$^{2}$, 
        R. Falomo$^{3}$, 
	D. Bettoni$^{3}$\\
	$^{1}$ Tuorla Observatory, Department of Physics and Astronomy, University of Turku, V\"ais\"al\"antie 20 , FI-21500 Kaarina, Finland.\\
       	$^{2}$ Finnish Centre for Astronomy with ESO (FINCA), University of Turku, V\"ais\"al\"antie 20, FI-21500 Kaarina, Finland\\
       	$^{3}$ INAF -- Osservatorio Astronomico di Padova, Vicolo dell'Osservatorio 5, I-35122 Padova (PD), Italy\\
	}

\maketitle

\label{firstpage}

\begin{abstract}

We study the environments of low redshift ($z < 0.5$) quasars based on a large and homogeneous dataset from the Stripe 82 region of the Sloan Digital Sky Survey (SDSS). We have compared the $< 1$ Mpc scale environments of 302 quasars that were resolved in our recent study to those of 288 inactive galaxies with closely matched redshifts. Crucially, the luminosities of the inactive galaxies and the quasar host galaxies are also closely matched, unlike in most previous studies.

The environmental overdensities were studied by measuring the number density of galaxies within a projected distance of 200 kpc to 1 Mpc. The galaxy number density of the quasar environments is comparable to that of the inactive galaxies with similar luminosities, both classes of objects showing significant excess compared to the background galaxy density for distances $<$ 400 kpc. There is no significant dependence of the galaxy number density on redshift, quasar or host galaxy luminosity, black hole mass or radio loudness. This suggests that the fueling and triggering of the nuclear activity is only weakly dependent on the local environment of quasars, and the quasar phase may be a short-lived common phase in the life cycle of all massive galaxies.

Key words: galaxies: active - galaxies:nuclei - quasars: general

\end{abstract}

\bigskip

\section{Introduction}

The last decades have seen the emergence of the general consensus that most, if not all, massive galaxies host a supermassive black hole in their center \citep[e.g.][]{richstone98}. Observations of early-type galaxies have shown a tight relation between the mass of the central black holes and the properties of the spheroids hosting them \citep[see e.g.][for a review]{ferrarese06} which suggests that the formation and evolution of the galaxies and their nuclear activity are linked. As quasars are fueled by accretion onto the supermassive black hole \citep[e.g.][]{yu02}, understanding the mechanism that triggers their activity plays a fundamental role in our understanding of the processes that have built the galaxies and their nuclei.

In spite of half a century of studies aimed to understand the quasar activity, the mechanism that activates and fuels the nuclei of galaxies is still a matter of debate. The leading processes thought to be responsible for transforming a dormant massive black hole into a luminous quasar are dissipative tidal interactions and galaxy mergers \citep[e.g.][and references therein]{dimatteo05,callegari11}. Galaxy formation is known to be heavily influenced by the environment, with galaxies in clusters tending to be elliptical and deprived of most of their gaseous content \citep[e.g.][]{silk93,kormendy09}, and also commonly showing signs of close interactions and mergers \citep[e.g.][]{bennert08,mcintosh08}. To better understand how quasars are formed, it is therefore important to study the relation between this environment and the nuclear activity.

The environments of quasars have been studied in the past on widely different scales ranging all the way from host galaxies to Mpc scales. Very early studies such as \citet{stockton78} and \citet{yee84} have shown that typical quasar environments have galaxy densities comparable to galaxy groups or poor clusters. At Mpc scales, comparing the environments of quasars to those of galaxies has given conflicting results. Early studies on Mpc scales suggest that quasars are more strongly clustered than galaxies \citep[e.g.][]{shanks88,chu88}, while later studies based on surveys such as the Two Degree Field (2dF) and the Sloan Digital Sky Survey (SDSS) have found the environmental galaxy densities of quasars and galaxies to be comparable to each other \citep[e.g.][]{smith00,wake04}.

At smaller scales, the studies have also shown differing results. \citet{ellingson91} studied a sample of 32 radio loud quasars (RLQs) and 33 radio quiet quasars (RQQs) at $0.3 < z < 0.6$ and found that the environments around RLQs are significantly denser than those around RQQs, which they find to have environmental density values similar to those found for average non-active galaxies in previous studies. More recent studies of small samples of low redshift quasars such as \citet{fisher96} and \citet{mclure01} on the other hand find no difference between the environments of RLQs and RQQs. Both studies used data taken with the HST, with \citet{fisher96} studying a sample of 20 quasars at $z \leq 0.3$, and \citet{mclure01}) using a sample of 44 quasars at $z ~0.2$. The environments of both RLQs and RQQs were found to have densities larger than those of non-active galaxies, with values similar to those found for the RLQ sample of \citet{fisher96}. Similar results were reported by \citet{wold00} and \citet{wold01} who used images from Nordic Optical Telescope (NOT) and HST to study 21 RLQs and 21 RQQs at $0.5 < z < 0.8$, concluding that no evolution of the environmental density with redshift was found. A more recent study by \citet{almeida13} compared the environments of 19 radio galaxies at redshifts $0.2 < z < 0.7$ to those of 20 RQQs at $0.3 < z < 0.41$ and found that radio galaxies appear to reside in denser environmets than quasars.

The early studies of quasar environments had to deal with small sample sizes, but surveys such as the Two Degree Field (2dF) and Sloan Digital Sky Survey (SDSS) have allowed studies with much larger quasar and galaxy samples. \citet{croom04}, for example, used data from the 2dF QSO Redshift Survey (2QZ) to study the clustering of of $~20000$ quasars at $z < 3$. The study of the actual environments of the quasars and a comparison sample of galaxies was done at redshifts $z < 0.3$ for a subsample of $~200$ quasars, and \citet{croom04} found the environments of the quasars to be statistically identical to those of galaxies. \citet{coldwell06} used the third data release (DR3) of SDSS to study the environments of $~2000$ quasars at redshifts of $z < 0.2$, using a comparison sample of 2300 galaxies. The quasar and galaxy samples were selected to have similar redshift distributions, but no matching based on the luminosity of the galaxies was done. The study found no difference between the environmental density of the two samples, noting that both quasars and galaxies tend to reside in regions more dense than field galaxies, but less dense than cluster environments. 

However, some other studies using the SDSS archives have found contradicting results. \citet{serber2006} also used data from SDSS DR3 to study environments of $~2000$ quasars at $z < 0.4$ and found again that quasars are located in regions of local overdensity higher than that of the background and that the density enhancement is strongest within 100 kpc from the quasar. They also found that the overdensity around the quasars is larger than that around typical L* galaxies, and that the high luminosity quasars have denser small-scale environments than low luminosity quasars. Another study by \citet{strand2008} used SDSS DR5 to study a sample of ~4000 quasars at $z < 0.6$, found environmental densities similar to those of \citet{serber2006}, though a comparison with non-active galaxies could not be performed due to the lack of a control sample. A more recent study by \citet{zhang13} used data from Stripe 82 region of the SDSS to study environments of 2300 quasars at $0.6 < z < 1.2$. They found that quasars exhibit an overdensity of galaxies with respect to the background of field galaxies, and that the clustering amplitude increases with the redshift. However, in the absence of a control sample, it is unclear whether the density of galaxies around quasars actually differs from the density around non-active galaxies at these redshifts.

In this study we have used more recent data from the seventh SDSS data release (DR7) \citep{abazajian09} to study the environments of low redshift quasars. In order to perform the environment study with deeper capability, we have used a stripe of sky along the Celestial Equator in the Southern Galactic Cap known as Stripe 82 \citep{annis2011}. This region was imaged multiple times during the period of 2004-2007, and the final co-added images reach up to 2 magnitudes deeper than other SDSS data. This allowed us to use a fainter magnitude limit for the galaxies to be included than the previous studies. 

This work is part of a series of papers aimed at investigating the properties of low redshift quasars from a large and homogeneous dataset. In paper I \citep{falomo2014} the results on the properties of quasar host and the relationship with BH mass were reported. In paper III (Bettoni et al. in prep) we aim to study the morphology, colours and peculiarities of quasar hosts, while paper IV will study the colours of galaxies in the environments of quasars. The work done in paper I allowed us to construct a control sample of galaxies which is closely matched with the quasar sample with respect to the properties of the host galaxies, an important aspect which was neglected by previous studies. The quasar and galaxy samples used in the study are described in Section 2. The dataset and the method of analysis are described in Section 3, and the results are given in Section 4. Section 5 contains comparison with previous works. Finally in Section 6 we report the main conclusion of this study.

We adopt the concordance cosmology with H$_0$ = 70 km s$^{-1}$ Mpc$^{-1}$, $\Omega_m$ = 0.3 and $\Omega_\Lambda$ = 0.7.

\section{The sample}

\begin{figure}
\centering
\includegraphics[width=\columnwidth]{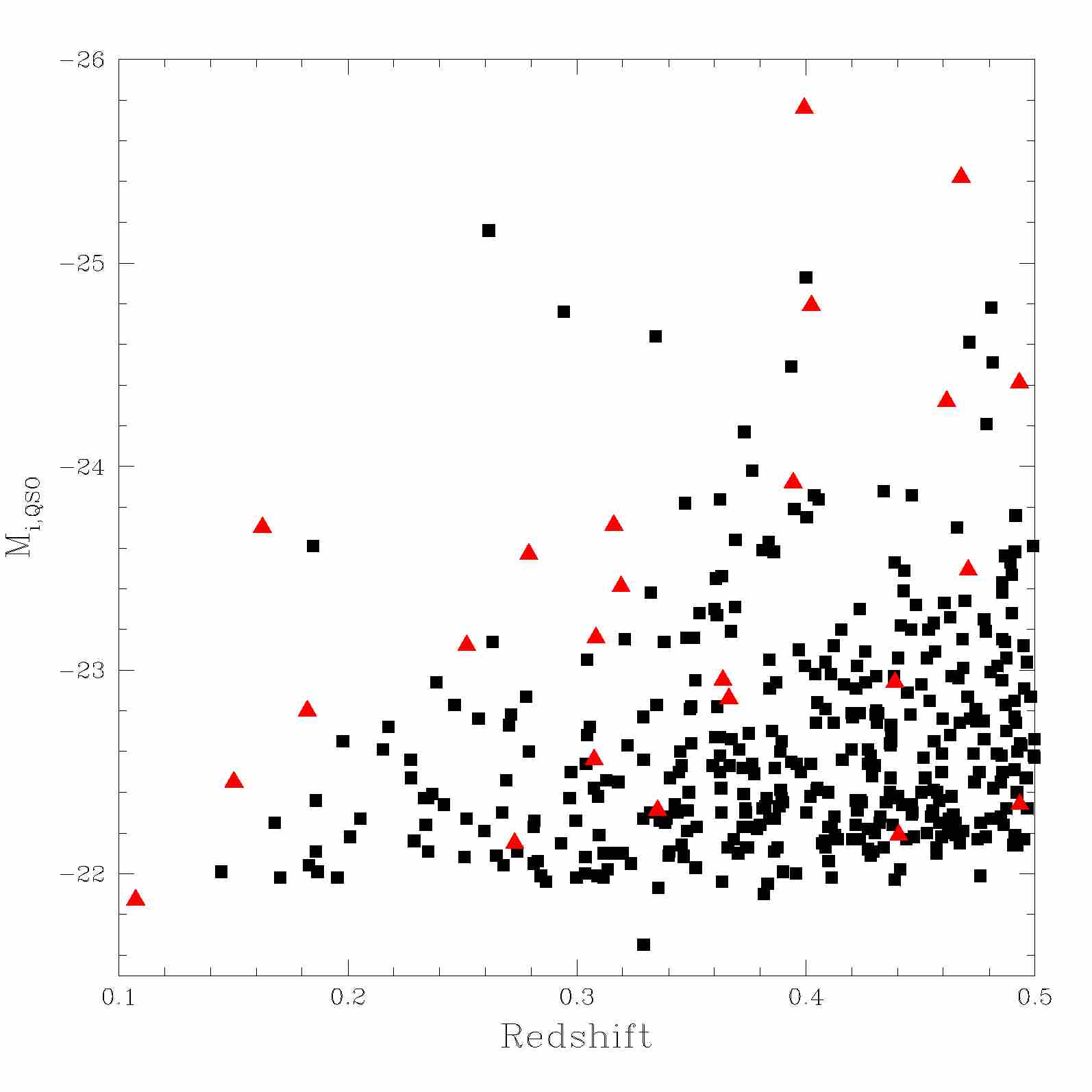}
\caption{Distribution of the FQS (416 objects) in z-$M_i$ plane. RQQs are represented by black squares and RLQs by red triangles.}
\label{fig:sampleq}
\end{figure}

\begin{figure*}
\centering
\includegraphics[width=\textwidth]{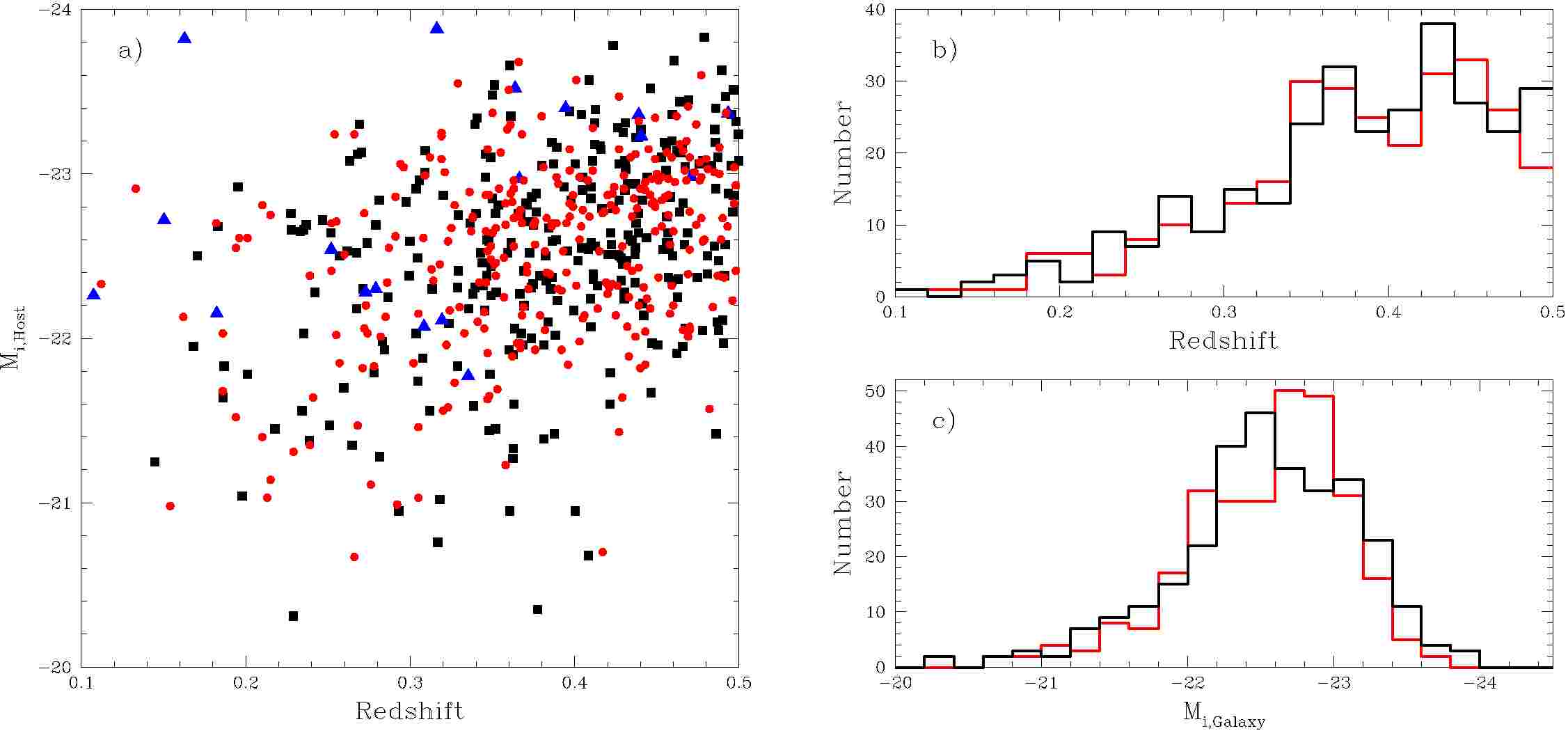}
\caption{The properties of the RQS and the MCGS, where the $M_i$ of the RQS refers to the magnitude of the host galaxy, as determined by Falomo et al. (2014): a) distribution of the samples in z-$M_i$ plane, RQQs are represented by black squares, RLQs by blue triangles and control sample galaxies by red circles; b) Distributions of the two samples with respect to redshift (black for RQS, red for MCGS); c) Distributions of the two samples with respect to $M_i$ of the (host) galaxies for both the RQS (black) and MCGS (red).}
\label{fig:sample}
\end{figure*}

The quasar sample was extracted from the fifth release of the SDSS Quasar Catalog \citep{schneider2010}, which uses data from the seventh SDSS data release \citep{abazajian09}. The catalog consists of quasars fainter than $i\sim15.0$, with an absolute magnitude of $M_i<$-22.0 and a reliably measured redshift. Additionally the quasars either have at least one emission line with FWHM$>$1000 km/ sec, or show complex/interesting absorption lines. The full catalog contains $\sim$106.000 spectroscopically confirmed quasars. In this work we only used objects in the Stripe 82 \citep{annis2011} region.

There were two main constraints imposed to obtain our final quasar sample. First, we had to avoid objects too close to the edges of the Stripe 82, to make sure we could fully study the environment of each standard object. Second, since the goal was to use the same quasar sample to study both the environments (this work) and the host galaxies \citep[see][]{falomo2014}, the high redshift limit had to be chosen such that resolving the quasar host galaxy would be possible for most targets. 

Our quasar sample includes quasars from the above catalog with a redshift in the range $0.1<z<0.5$, which are inside the region $1.0<DEC<-1.0$, $0<RA<59.8$ and $300.2<RA<360$. This gives a total of 416 quasars. Of these, 24 quasars were detected by the FIRST radio survey \citep{becker95}, with a detection threshold of 1 mJy. These quasars form the radio loud quasar (RLQ) subsample. The mean redshift of the whole quasar sample is $<z>$ =$0.39\pm0.08$ and the average absolute magnitude is $<M_i>$ = $-22.68\pm0.62$. Fig. \ref{fig:sampleq} shows the distribution of this "full quasar sample" (FQS) in the z-$M_i$ plane, with the RLQs shown as red triangles. As we were also interested in studying the magnitudes of the quasar host galaxies and their possible dependence on the environment, we also defined a smaller sample containing the objects resolved by \citet{falomo2014}. This "resolved quasar sample" (RQS) contains 302 objects and has $<z>$ =$0.38\pm0.08$, with $<M_{i,qso}>$ = $-22.30\pm0.77$ and $<M_{i,host}>$ = $-22.54\pm0.63$.

In order to see how the environments of quasars compare to those of inactive galaxies, we also defined a control sample of galaxies with similar redshift and host galaxy magnitude distribution. To do this, we selected all objects in the Stripe 82 database that are classified as galaxies for which accurate spectroscopic redshifts had been determined. The sample was further cut to have redshifts in the range $0.1<z<0.5$. Finally, a subset of 580 galaxies was chosen with a redshift distribution close to that of the FQS. This "full control galaxy sample" (FCGS) has a mean redshift of $<z>$ =$0.38\pm0.08$ and an average absolute magnitude of $<M_i>$ = $-22.22\pm0.77$. This sample was compared with the host galaxy magnitudes of the RQS, determined by \citet{falomo2014} to make the magnitude distribution of the samples similar. Our final "matched control galaxy sample" (MCGS) contains 288 objects, with a mean redshift of $<z>$ =$0.38\pm0.08$ and an average absolute magnitude of $<M_R>$ = $-22.53\pm0.55$. The distribution of the objects in the z-$M_i$ plane for the RQS and MCGS are shown in Fig. \ref{fig:sample}a. Fig. \ref{fig:sample}b and \ref{fig:sample}c show the comparison of the redshift and (host) galaxy magnitude distributions for the two samples.

\section{Data and Analysis}

\begin{figure*}
\centering
\includegraphics[width=\textwidth]{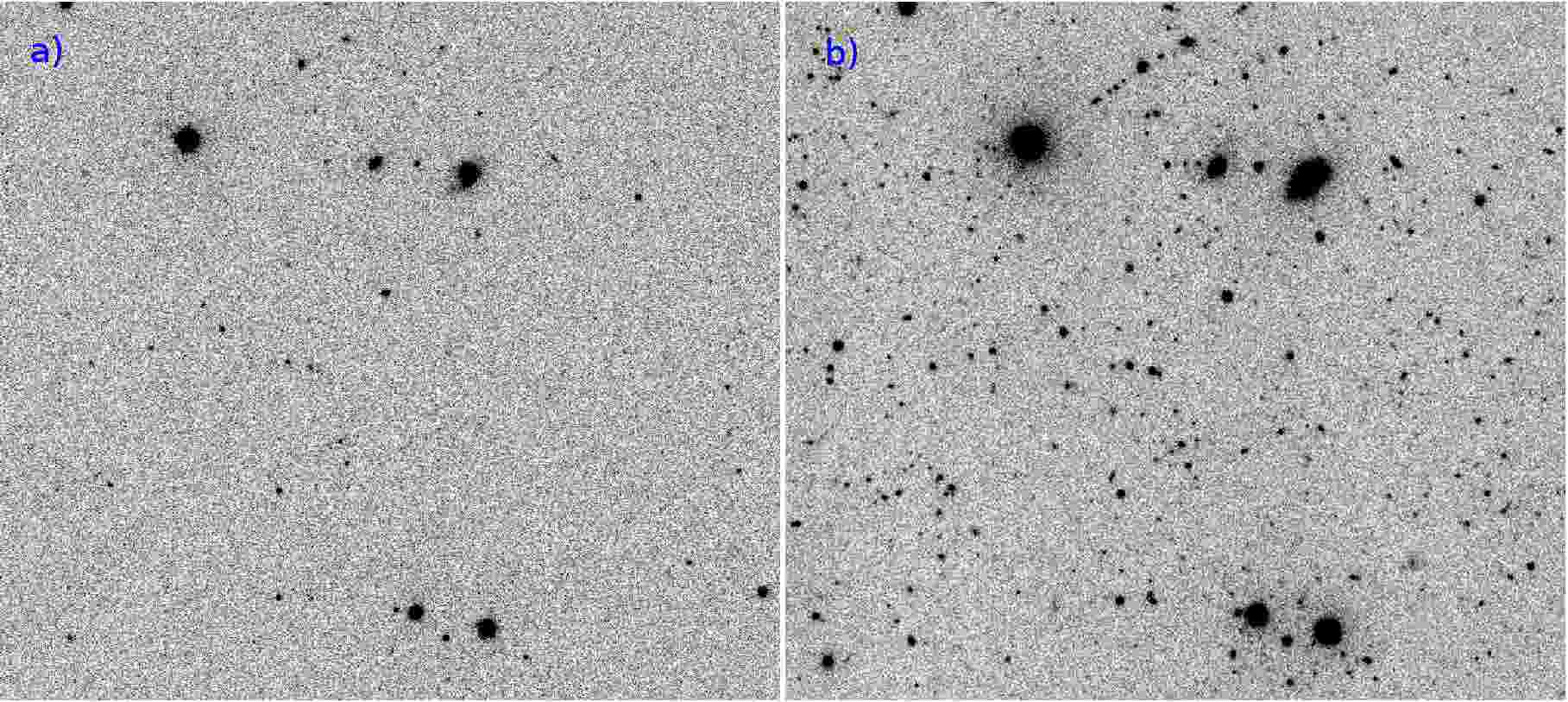}
\caption{An example showing the difference between the i-band images in SDSS DR7 (panel a) and Stripe 82 (panel b) datasets for the region around SDSS J023922.87-000119.5. This Stripe 82 image is a combination of 67 individual images.}
\label{fig:sdsscomp}
\end{figure*}

We used the i-band images to study the quasar environments. These images correspond to the R filter at the rest frame of an object at the average redshift of the dataset. The advantage of using images of the Stripe 82 is illustrated in Fig. \ref{fig:sdsscomp}, which shows an image from SDSS DR7 compared with the same area from the Stripe 82 data.

SDSS archives provide a catalog of photometric objects in Stripe 82, classified into galaxies and stars. This catalog provides positions and magnitudes of the objects. In order to have full control of the measurements we performed independent detection and classification of objects in the images using software SExtractor \citep{bertin96}. We compared the objects catalogs generated with SExtractor to those of SDSS, and performed a visual inspection on a number of frames to further to study the validity of our classification. We found a good match between the SDSS and SExtractor catalogs at apparent magnitudes $m_i < 23$, but at fainter magnitudes the number of objects detected by SExtractor dropped dramatically compared to those in the SDSS catalog. A visual inspection of these faint objects showed that they are mainly background noise which is either undetected by SExtractor or classified as an "unknown" object. (See Appendix A for details)

The Stripe 82 region has a significant trend in the stellar density, caused by differences in the Galactic latitude over the stripe, ranging from $-25^{\circ} < b < -65^{\circ}$. As a result, up to 50\% of objects in the fields located at low Galactic latitudes are stars. At higher Galactic latitudes, the fraction of objects classified as stars dropped down to $\sim$10\%. The large number of stars in some of our fields thus required us to have a robust method for separating stars and galaxies.

\begin{figure}
\centering
\includegraphics[width=\columnwidth]{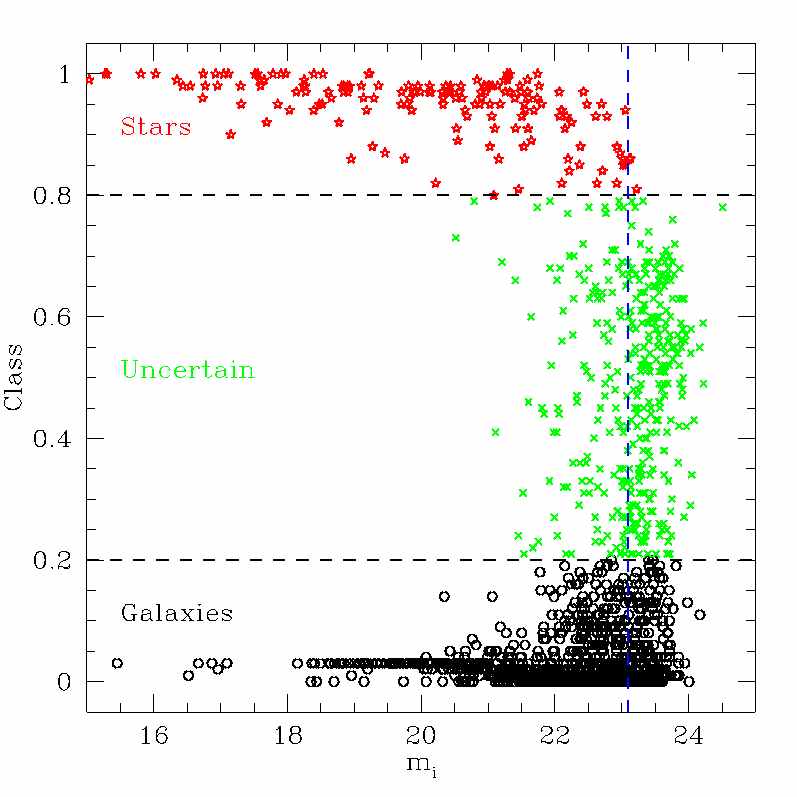}
\caption{A plot of Star-Galaxy classification vs Magnitude for the field containg SDSS J021447.00-003250.6. Black circles show the galaxies, while red stars and green crosses denote stars and objects with uncertain classification, respectively. The blue dashed line shows the magnitude threshold determined for the frame.}
\label{fig:obj311cl}
\end{figure}

The object detection and classification in the images was done by SExtractor \citep{bertin96}, which uses a neural network to determine how star- or galaxy-like an object is. SExtractor was selected because of its fast processing time, and the ability to classify faint objects more accurately than other similar catalog extraction softwares \citep[e.g.][]{becker07,annunz13}. This classification is represented as a number from 0.0 to 1.0, with 0.0 corresponding to a galaxy and 1.0 corresponding to a star. Objects for which classification is uncertain are given values between these two extremes. To minimise the number of misclassified objects in our analysis, we studied the distribution of the classification versus the magnitude, an example of which is shown in Fig. \ref{fig:obj311cl}. We chose a conservative limit of classification value $\leq$ 0.20 for our galaxies, to make sure we avoid all the stars and majority of the "unknown" objects (centered around 0.5). Furthermore, the majority of the "unknown" objects are fainter than the magnitude thresholds we calculated for the frames (see below), thus making the selection of the exact galaxy limit less sensitive.

\begin{figure*}
\centering
\includegraphics[width=\textwidth]{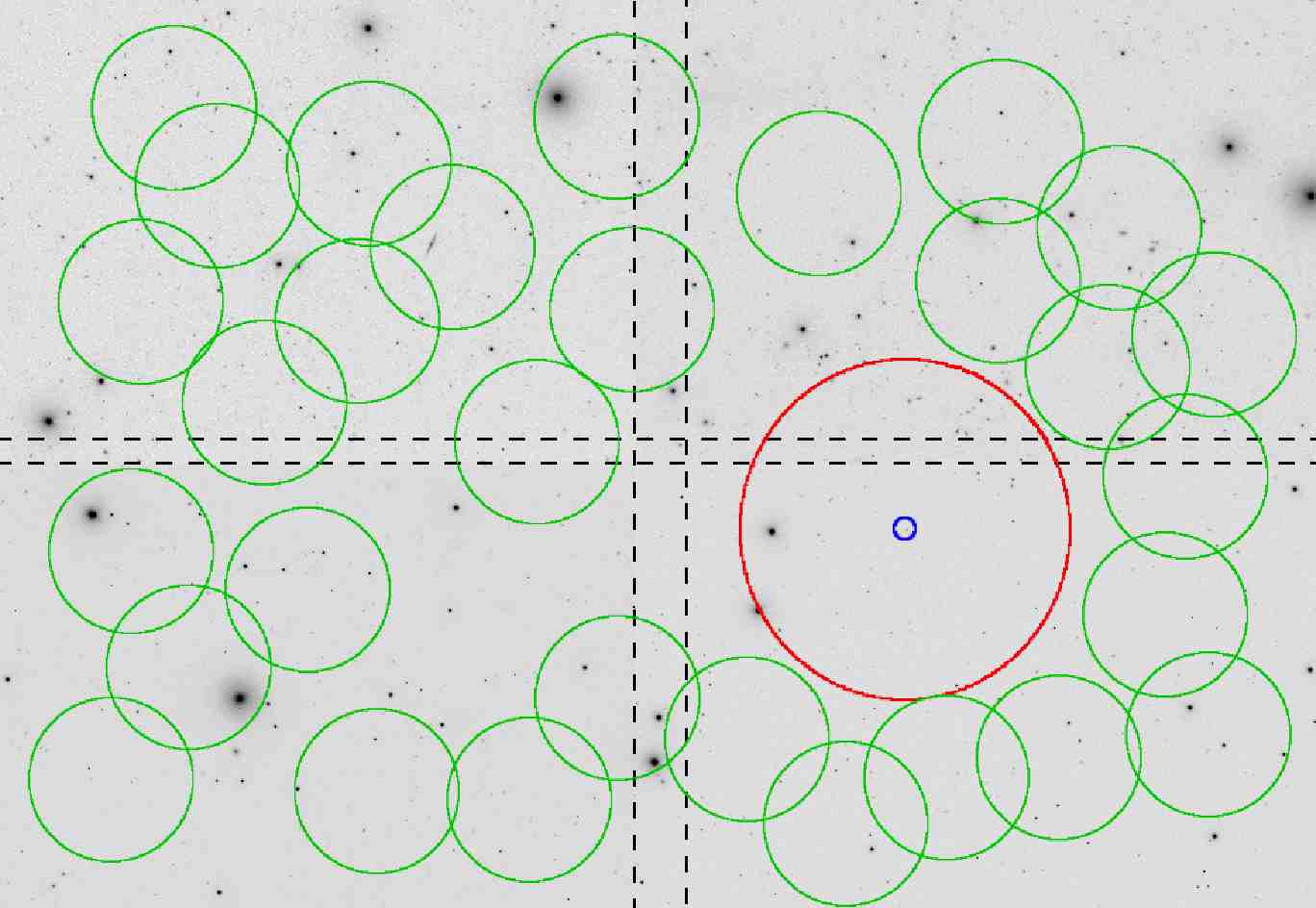}
\caption{The four frames used to study the environment of SDSS J021447.00-003250.6. The redshift of the object is 0.3489. The dashed black lines show the edges of each of the 4 overlapping frames. The blue circle shows the location of the target, while the red circle corresponds to a distance of 1 Mpc from the target. The green circles show the randomly picked areas used for determining the background galaxy density.}
\label{fig:obj311ff}
\end{figure*}

In addition to the frames containing the targets in our quasar and control samples, we also considered three additional adjacent frames located closest to the targets. This was done because the frames were not in general centered around our target objects, so the environment up to a distance of 1 Mpc from the target often didn't fit within the frame. This is illustrated in Fig. \ref{fig:obj311ff}, where the frame containing the target quasar is located in lower right corner. Additionally, at low redshifts, the field of view of the SDSS images (13.5\arcmin $\times$ 9.8\arcmin) is too small to fit the full environment up to a distance of 1 Mpc; for example, at redshift z = 0.17, a radius of 1 Mpc corresponds to 5.8 arcminutes. A larger field of view is also required to determine the background galaxy density.

The input parameters for Source Extractor were determined separately for each analysed image, to account for the fact that the image quality may be different from one frame to another. The effect of artifacts in the images was minimized by excluding from the analysis the objects flagged by Source Extractor as having close, bright neighbours, with which they had been originally blended. This also removed some genuine objects, but the loss was generally $\sim5\%$ of total number of detected objects. Once the Source Extractor had been run on the "target frame" and the three associated "environment frames" the four object catalogs were combined into one "master" catalog. As the 4 frames overlapped with each other on the edges, some of the objects in the overlapping areas were essentially detected twice. This was dealt with by using the detected objects from only one of the frames in the overlapping regions, giving preference to the target frame.

\begin{figure}
\centering
\includegraphics[width=\columnwidth]{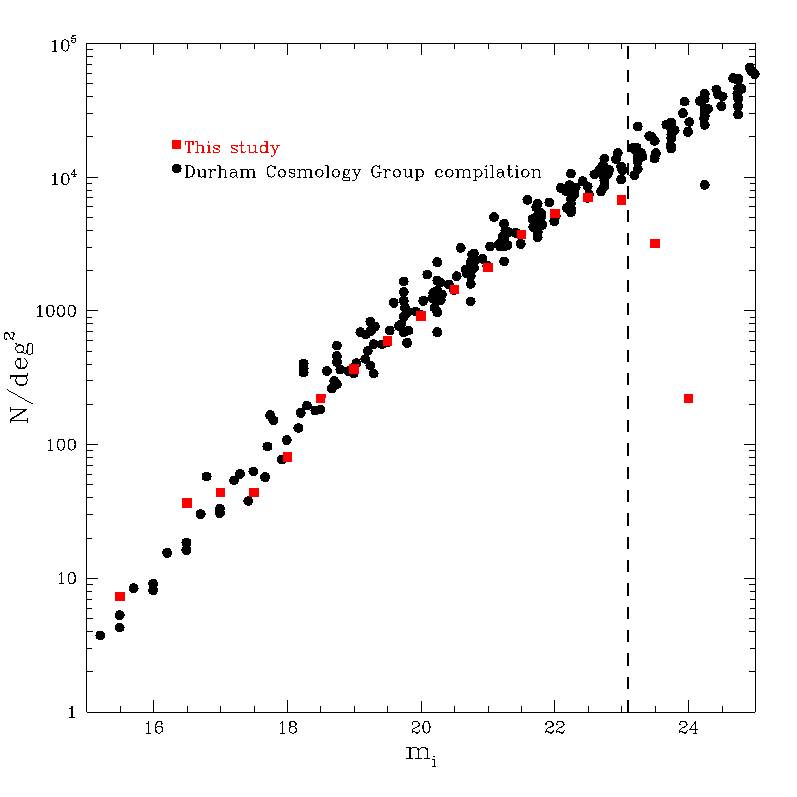}
\caption{A plot of galaxy counts vs. $m_i$ for both the field containing SDSS J021447.00-003250.6 (red squares) and the data from Durham Cosmology Group (black dots). The dashed black line shows the magnitude limit of the field.}
\label{fig:obj311gal}
\end{figure}

\begin{figure}
\centering
\includegraphics[width=\columnwidth]{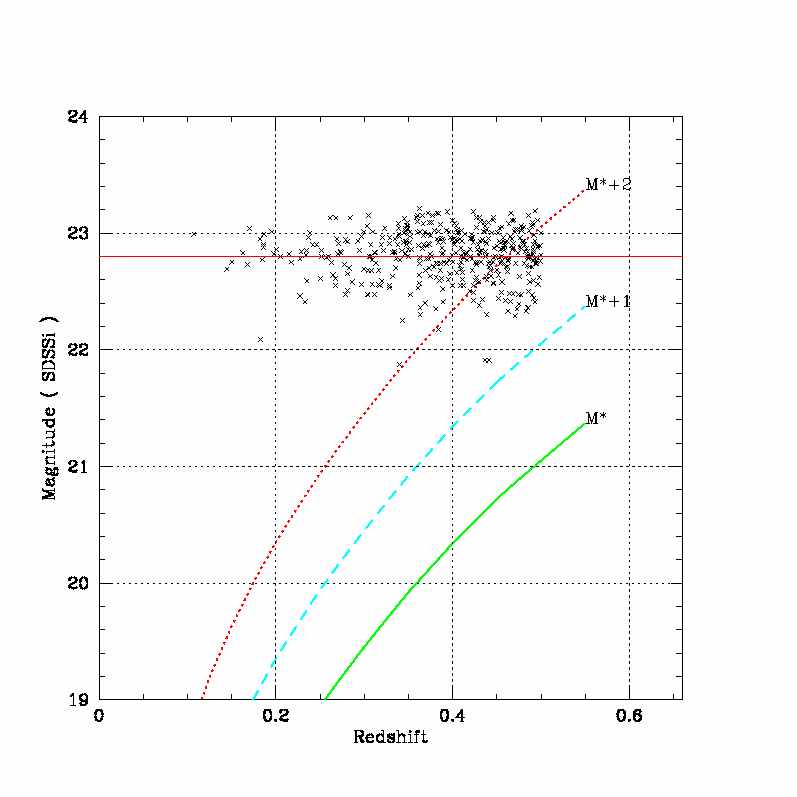}
\caption{A plot of derived magnitude thresholds in $m_i$ vs. the redshift of the object. The solid black line shows the mean value of the threshold for our sample, $m_i = 22.8$. The solid green line shows the apparent i-band magnitude of M* at different redshifts, with the dashed cyan and dotted red lines corresponding to M*+1 and M*+2, respectively.}
\label{fig:magthresh}
\end{figure}

To determine the magnitude threshold of our frames, we compared the galaxy number counts in each of the fields to those expected by the Extragalactic Astronomy \& Cosmology Research Group at Durham University\footnote{Data and references available online at \url{http://astro.dur.ac.uk/~nm/pubhtml/counts/counts.html}}. The magnitude threshold for each frame was then defined as the magnitude at which completeness in the image had dropped to 50\%. As each of the 4 frames used to study one target could be of slightly different quality, and hence had a different magnitude threshold, we chose to set our limits according to the brightest magnitude threshold. Fig \ref{fig:obj311gal} shows a comparison between the galaxy counts in the field of one target, and the data from the Durham Cosmology Group. The dashed line shows the magnitude limit adopted for the frame. The average value of the magnitude threshold for the FQS is $m_i = 22.8$, and the distribution of the magnitude thresholds is shown in Fig. \ref{fig:magthresh}, along with M*, M*+1 and M*+2 at different redshifts, where M*(i) = -21.9 \citep{loveday12}. It is noticeable that for the majority of target the magnitude threshold allows us to measure galaxies as faint as M*+2.

\subsection{Galaxy background determination}

The background galaxy density was determined for each object by selecting randomly 30 sets of positions within the area covered by the combined field. The number density of galaxies within a 250 pixel radius (corresponding to 99 arcsec or a physical radius of $\sim$190-600 kpc at our redshift range) was then calculated for each of these positions. These random positions had a minimum distance of 250 pixels from the edges of the field, and the coordinates were selected in way that none of the background areas overlapped with the area around the quasar. This procedure is illustrated in Fig. \ref{fig:obj311ff}. The value for the background density was determined by taking the average value the galaxy density from of all the positions.

\section{Galaxy environment of quasars}

\begin{figure*}
\centering
\includegraphics[width=\textwidth]{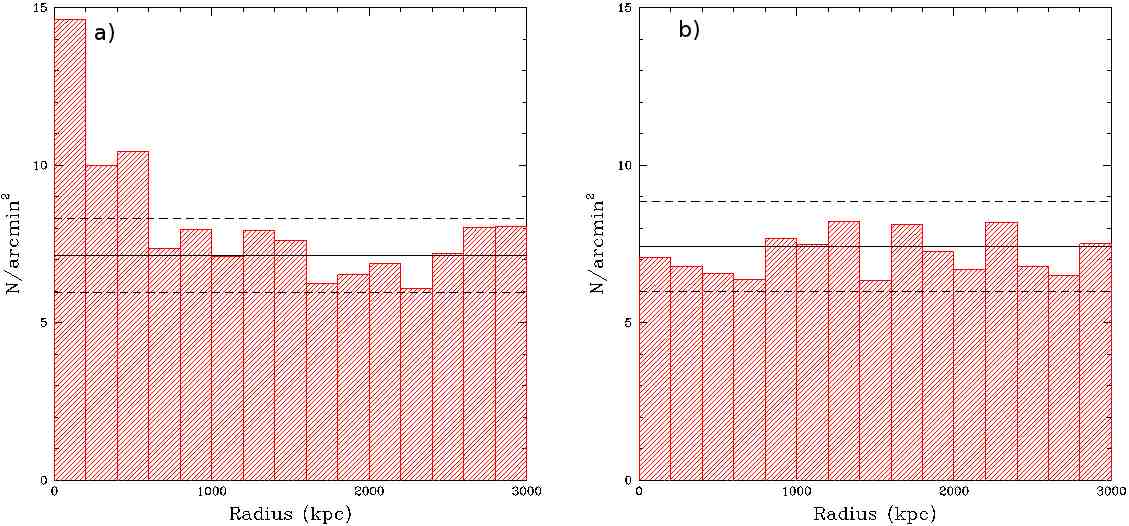}
\caption{An example of radial distribution of galaxy surface number density around two quasars up to a distance of 3 Mpc: a)SDSS J021447.00-003250.6 at a redshift of z = 0.35, b) SDSS J003723.49+000812.5 at a redshift of z = 0.25. The solid black line shows the background galaxy density of the field with it's associated uncertainties (dashed black lines).}
\label{fig:objdens}
\end{figure*}

The projected galaxy number densities around the quasars were determined by counting the number of detected galaxies within a radius $R_{max}$, and then dividing the number by the (angular) area covered. This was done for different distances from 100 kpc to a maximum distance of 3 Mpc. Tables \ref{tab:bgqso} and \ref{tab:bgcon} show the number of galaxies and the calculated galaxy density within a distance of 250kpc for the first 20 objects of the FQS and FCGS, respectively. In addition to this cumulative method, the environment was also divided into annular regions with a fixed width of 200 kpc at the redshift of the target, and the number density of galaxies in each of these radius bins out to a distance of 3 Mpc was derived. The densities computed by the latter method are illustrated in Fig. \ref{fig:objdens} for two representative objects. In one case a clear increase in density is observed at small radii while for the other the density is at the same level as the background.

\begin{table*}
\caption{The measured galaxy number densities within a distance of 0.25 Mpc, background galaxy number densities and magnitude limits for the FQS. $N_{0.25}$ is the number of galaxies detected within 0.25 Mpc of the quasar, {$n_{0.25}$} is the galaxy number density within 0.25Mpc, $n_{bg}$ is the average background galaxy density and $m_{i,th}$ is the apparent magnitude threshold determined for the frame. Only the first 20 quasars are  shown. The complete table is available in electronic format.}
\begin{tabular}{|l|l|r|r|r|r|r|r|}
\hline
  \multicolumn{1}{|c|}{Nr$^{a}$} &
  \multicolumn{1}{c|}{SDSS} &
  \multicolumn{1}{c|}{z} &
  \multicolumn{1}{c|}{$N_{0.25}$} &
  \multicolumn{1}{c|}{$n_{0.25}$} &
  \multicolumn{1}{c|}{$n_{bg}$} &
  \multicolumn{1}{c|}{$\sigma_{bg}$} &
  \multicolumn{1}{c|}{$m_{i,th}$}\\
& & & arcmin$^{-2}$ & arcmin$^{-2}$ & arcmin$^{-2}$ & mag \\
\hline
1 $^{r}$ & J203657.28+000144.3 & 0.44 & 5 & 2.97 & 3.12 & 0.70 & 21.91\\
2 & J203746.78+001837.2 & 0.45 & 12 & 7.30 & 5.28 & 1.13 & 22.68\\
3 & J203905.23-005004.9 & 0.43 & 10 & 5.72 & 4.84 & 1.21 & 22.62\\
4 $^{r}$ & J204153.51+002909.8 & 0.40 & 10 & 5.24 & 4.15 & 0.83 & 22.65\\
5 $^{r}$ & J204340.03+002853.4 & 0.32 & 18 & 7.05 & 5.97 & 1.31 & 22.68\\
6 & J204433.61+005035.5 & 0.49 & 6 & 3.97 & 3.99 & 1.20 & 22.36\\
7 & J204527.70-003236.2 & 0.30 & 18 & 6.45 & 3.96 & 0.79 & 22.56\\
8 $^{r}$ & J204621.29+004427.8 & 0.40 & 5 & 2.65 & 3.61 & 1.00 & 22.55\\
9 & J204626.10+002337.7 & 0.33 & 10 & 4.18 & 4.42 & 0.94 & 22.63\\
10 $^{r,q}$ & J204635.37+001351.7 & 0.49 & 10 & 6.62 & 4.63 & 0.72 & 22.65\\
11 $^{r}$ & J204753.67+005324.0 & 0.36 & 6 & 2.82 & 3.05 & 0.86 & 22.36\\
12 $^{r}$ & J204826.79+005737.7 & 0.49 & 6 & 3.97 & 3.29 & 0.86 & 22.36\\
13 $^{r}$ & J204844.19-004721.5 & 0.47 & 4 & 2.53 & 4.31 & 0.90 & 22.41\\
14 & J204910.96+001557.2 & 0.36 & 10 & 4.69 & 5.56 & 0.89 & 22.79\\
15 & J204936.47+005004.6 & 0.48 & 5 & 3.23 & 3.85 & 0.84 & 22.39\\
16 $^{r,q}$ & J204956.61-001201.7 & 0.37 & 11 & 5.27 & 4.28 & 0.84 & 22.68\\
17 $^{r}$ & J205050.78+001159.7 & 0.31 & 22 & 8.33 & 5.39 & 0.93 & 22.77\\
18 $^{r}$ & J205105.02-003302.7 & 0.30 & 14 & 5.09 & 5.14 & 0.97 & 22.69\\
19 $^{r}$ & J205212.28-002645.2 & 0.27 & 11 & 3.40 & 4.91 & 0.94 & 22.61\\
20 $^{r}$ & J205352.03-001601.5 & 0.36 & 10 & 4.68 & 5.23 & 1.19 & 22.74\\
\hline
\end{tabular}
\begin{list}{}
\item   $^{(a)}$ Quasars belonging to RQS indicated by $^{(r)}$, while quasars with $^{(q)}$ and $^{(l)}$ belong to the RQQ and RLQ samples, respectively\\
\end{list}
\label{tab:bgqso}  
\end{table*}

\begin{table*}
\caption{The measured galaxy number densities within a distance of 0.25 Mpc, background galaxy number densities and magnitude limits for the FCGS. $N_{0.25}$ is the number of galaxies detected within 0.25 Mpc of the galaxy, {$n_{0.25}$} is the galaxy number density within 0.25Mpc, $n_{bg}$ is the average background galaxy density and $m_{i,th}$ is the apparent magnitude threshold determined for the frame. Only the first 20 galaxies are  shown. The complete table is available in electronic format.}
\begin{tabular}{|l|l|r|r|r|r|r|r|}
\hline
  \multicolumn{1}{|c|}{Nr$^{a}$} &
  \multicolumn{1}{c|}{SDSS} &
  \multicolumn{1}{c|}{z} &
  \multicolumn{1}{c|}{$N_{0.25}$} &
  \multicolumn{1}{c|}{$n_{0.25}$} &
  \multicolumn{1}{c|}{$n_{bg}$} &
  \multicolumn{1}{c|}{$\sigma_{bg}$} &
  \multicolumn{1}{c|}{$m_{i,th}$}\\
& & & arcmin$^{-2}$ & arcmin$^{-2}$ & arcmin$^{-2}$ & mag \\
\hline
1 $^{m}$ & J000121.46-001140.3 & 0.46 & 12 & 7.52 & 5.58 & 1.52 & 22.93\\
2 $^{m}$ & J000127.46+002815.4 & 0.25 & 23 & 6.60 & 6.71 & 1.56 & 22.80\\
3 & J000145.00+001900.9 & 0.46 & 10 & 6.25 & 6.30 & 1.25 & 22.84\\
4 $^{m}$ & J000152.82+003533.0 & 0.39 & 21 & 10.64 & 6.37 & 1.37 & 22.76\\
5 $^{m}$ & J000226.85+004533.5 & 0.38 & 13 & 6.46 & 5.23 & 1.52 & 22.48\\
6 & J000322.05+002255.0 & 0.21 & 28 & 6.21 & 6.60 & 1.30 & 22.88\\
7 & J000358.79-001252.0 & 0.36 & 28 & 13.08 & 6.59 & 1.49 & 22.97\\
8 $^{m}$ & J000511.42+005820.4 & 0.36 & 12 & 5.63 & 4.44 & 0.87 & 22.46\\
9 & J000532.27-002259.1 & 0.35 & 14 & 6.22 & 6.31 & 1.36 & 22.81\\
10 & J000534.96-005640.2 & 0.37 & 8 & 3.84 & 5.58 & 1.02 & 22.73\\
11 $^{m}$ & J000537.94+001404.4 & 0.32 & 19 & 7.52 & 7.00 & 1.56 & 23.00\\
12 $^{m}$ & J000721.83-005609.9 & 0.38 & 4 & 2.01 & 6.90 & 1.90 & 22.91\\
13 & J000737.68-005357.1 & 0.44 & 11 & 6.52 & 6.26 & 1.22 & 22.91\\
14 & J000842.26+001403.0 & 0.45 & 19 & 11.64 & 6.35 & 1.75 & 22.89\\
15 & J000856.14+001758.2 & 0.36 & 26 & 12.11 & 7.06 & 1.75 & 22.89\\
16 & J000942.92-003048.3 & 0.39 & 11 & 5.59 & 5.07 & 1.09 & 22.88\\
17 $^{m}$ & J001016.62-002305.4 & 0.21 & 23 & 4.95 & 6.02 & 1.10 & 22.93\\
18 & J001024.80+000724.7 & 0.24 & 36 & 9.68 & 7.39 & 1.37 & 22.99\\
19 $^{m}$ & J001029.09+004904.2 & 0.35 & 10 & 4.39 & 5.38 & 1.10 & 22.61\\
20 $^{m}$ & J001111.95+001557.5 & 0.42 & 13 & 7.34 & 7.65 & 1.24 & 23.06\\
\hline
\end{tabular}
\begin{list}{}
\item   $^{(a)}$ Galaxies belonging to MCGS indicated by $^{(m)}$ \\
\end{list}
\label{tab:bgcon}    
\end{table*}

\begin{figure}
\centering
\includegraphics[width=\columnwidth]{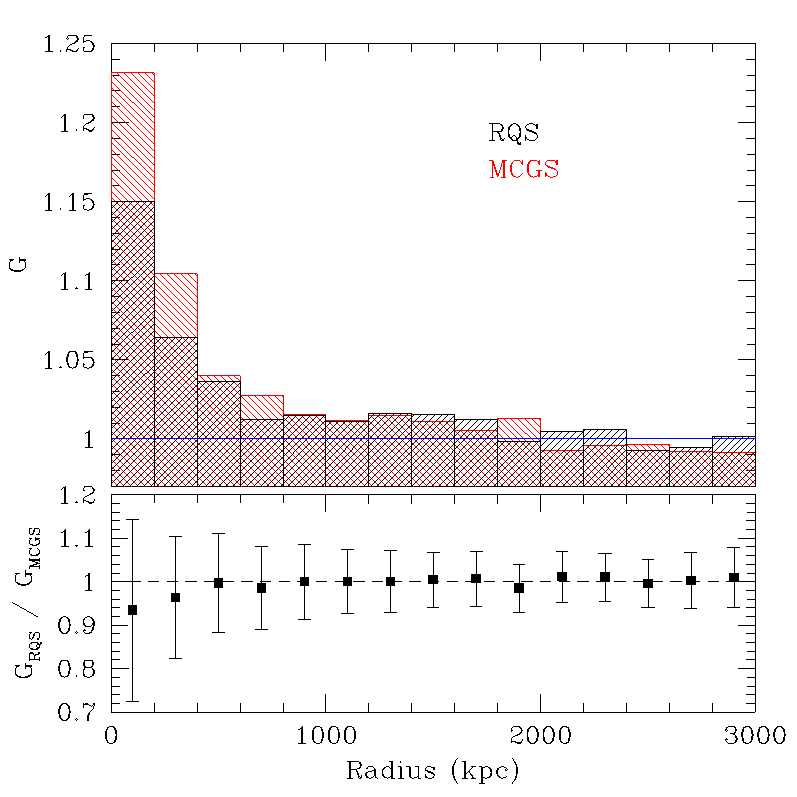}
\caption{Top Panel: Average differential overdensity of the environments for different distances from the target for both the RQS (black) and the MCGS (red). The solid blue line corresponds to no overdensity. Bottom panel: The ratio of the overdensities of the RQS and MCGS, with their associated uncertainties. The dashed line shows the case of equal overdensities.}
\label{fig:radovdavgcom}
\end{figure}

The excess galaxy surface density around the target, or the "overdensity", was defined as the ratio of the galaxy density measured near the target, $n_{env}$ and the background density, $n_{bg}$. We characterize this overdensity with a parameter G, where $ G = n_{env}/n_{bg}$. A ratio of unity (ie. $G = 1$) thus implies that no overdensity is present in the target environment. The overdensity was calculated for each of the 200 kpc wide radius bins surrounding the quasar to study how the density changes with projected distance from the quasar. The top panel of Fig. \ref{fig:radovdavgcom} shows the average differential overdensities for each bin for both the RQS and MCGS, while the bottom panel shows the ratio of the overdensities, $G_{RQS}/G_{MCGS}$. For both samples, the overdensity has a strong peak at the smallest separations (smaller than 200 kpc), and then quickly drops to the background level.  The average overdensity within 200 kpc reaches $1.15\pm0.04$ and $1.23\pm0.06$ for the RQS and MCGS, respectively, where the uncertainty represents the 95\% confidence level. The overdensity reaches values comparable to the background at about 1 Mpc for both samples. The overdensities at radii up to 1 Mpc are slightly larger for the control sample, but the difference isn't statistifically significant. This implies that the local environments of the quasars do not differ significantly from those of the galaxies in our control sample and in any case are not richer than those of inactive galaxies. As the overdensity at distances larger than 1 Mpc is similar to background, we have only performed the rest of the analysis at distances $<$ 1 Mpc.

\begin{figure*}
\centering
\includegraphics[width=\textwidth]{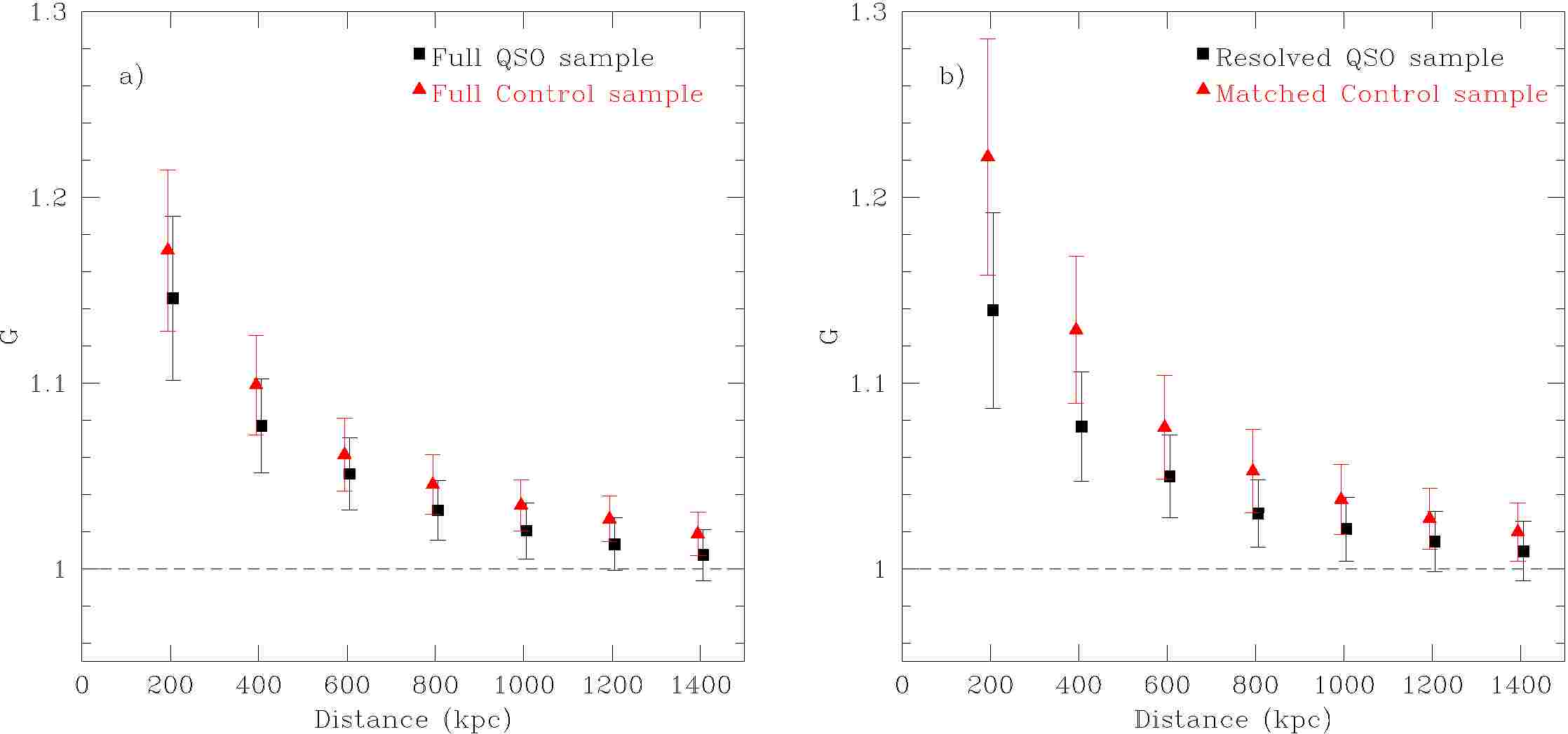}
\caption{Mean cumulative overdensity of the environments as a function of distance for a) the full quasar and control samples, b) the resolved quasar sample and the matched control sample. Black squares show the overdensity for the quasars and red triangles for the galaxies in the control sample. For clarity, points have been offset slightly horizontally.}
\label{fig:radovdcom}
\end{figure*}

Fig. \ref{fig:radovdcom} shows the mean cumulative overdensity around the targets in the RQS and MCGS. Table \ref{tab:ovden} shows the overdensities within different radii for the full and matched samples. The difference between the FQS and RQS is negligible, while the MCGS has larger overdensities than the FCGS, though the difference is not statistically significant; for example, at distance of 250 kpc, the FCGS shows an overdensity of $1.16\pm0.04$, while for the MCGS the same value is $1.21\pm0.06$. This small difference can be explained by the fact that the average luminosity of the galaxies in the MCGS is higher than that for the FCGS, and higher luminosity galaxies appear to have greater associated environmental overdensities (see section 4.3).

The highest excess densities are found at the small distances from the target, with the RQS having an excess of $1.22\pm0.10$ at 100 kpc radius, and the MCGS with an overdensity of $1.40\pm0.11$ at the same distance. The overdensity around the MCGS galaxies at larger radii is not significantly different from the overdensities found for the RQS.

Since the galaxy clustering around quasars could depend on a number of properties such as the redshift, nuclear luminosity, radio loudness and BH mass we search for possible correlations of the galaxy overdensity with these parameters.

\begin{table*}
\caption{The overdensities within a distance of $R_{max}$ around both our quasar and control samples at varying distances. The results from Table 1 of Serber et al. 2006 are also included for comparison.}
\begin{tabular}{|c|c|c|c|c|c|c|}
  \hline
  \multicolumn{1}{|c|}{$R_{max}$ (kpc)} &
  \multicolumn{2}{c|}{Quasar sample} &
  \multicolumn{2}{c|}{Control sample} &
  \multicolumn{2}{c|}{Serber et al. 2006}\\
  & FQS & RQS & FCGS & MCGS & Quasar & L* Galaxies\\
  \hline
  100 & $1.21 \pm 0.09$ & $1.22 \pm 0.10$ & $1.30 \pm 0.08$ & $1.40 \pm 0.11$ & $2.12 \pm 0.08$ & $1.507 \pm 0.010$\\
  250 & $1.11 \pm 0.04$ & $1.12 \pm 0.04$ & $1.16 \pm 0.04$ & $1.21 \pm 0.06$ & $1.57 \pm 0.03$ & $1.235 \pm 0.004$\\
  500 & $1.06 \pm 0.02$ & $1.06 \pm 0.03$ & $1.08 \pm 0.02$ & $1.10 \pm 0.03$ & $1.27 \pm 0.02$ & $1.144 \pm 0.003$\\
  1000 & $1.02 \pm 0.02$ & $1.02 \pm 0.02$ & $1.03 \pm 0.01$ & $1.04 \pm 0.02$ & $1.13 \pm 0.01$ & $1.082 \pm 0.002$\\
  \hline
\end{tabular}
\label{tab:ovden}
\end{table*}

\subsection{Dependence of environment on redshift}

\begin{figure}
\centering
\includegraphics[width=\columnwidth]{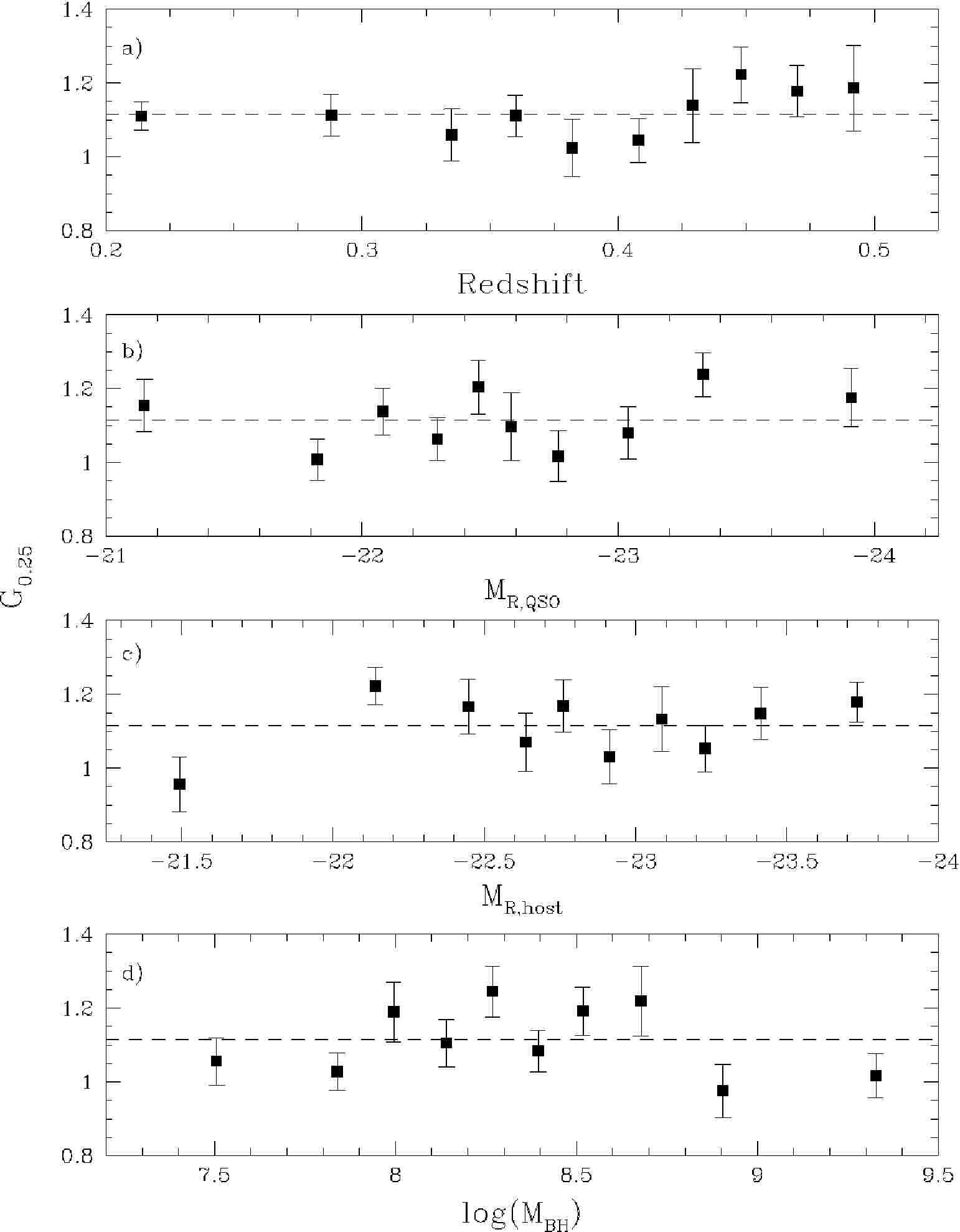} 
\caption{The dependence of the G$_{0.25}$ parameter on a) redshift, b) absolute magnitude of the quasar, c) absolute magnitude of the host galaxy and d) black hole mass of the quasar host. The dashed black line shows the mean value of G$_{0.25}$ for the quasar sample. In all cases the bins have been chosen to have the same number objects.}
\label{fig:g25depq}
\end{figure}

To study the possible evolution of the environments with redshift, we divided the RQS and MCGS into 10 bins, with the same number of objects in each redshift range. We then defined a parameter G$_{0.25}$ as the overdensity within a distance of 0.25 Mpc from the target to study the redshift dependence. We did a linear weighted least squares fit to the data to look for a dependence on redshift, and compared it with a zero slope fit for each sample. The fit parameters and their associated uncertainties are shown in Table \ref{tab:lsq} along with the probability of the fit, $P(\chi^2)$. Fig. \ref{fig:g25depq}a shows the G$_{0.25}$ parameter as a function of redshift for the RQS. The best linear fit has a slope of 0.15 $\pm$ 0.22 showing no redshift dependence.

\begin{figure}
\centering
\includegraphics[width=\columnwidth]{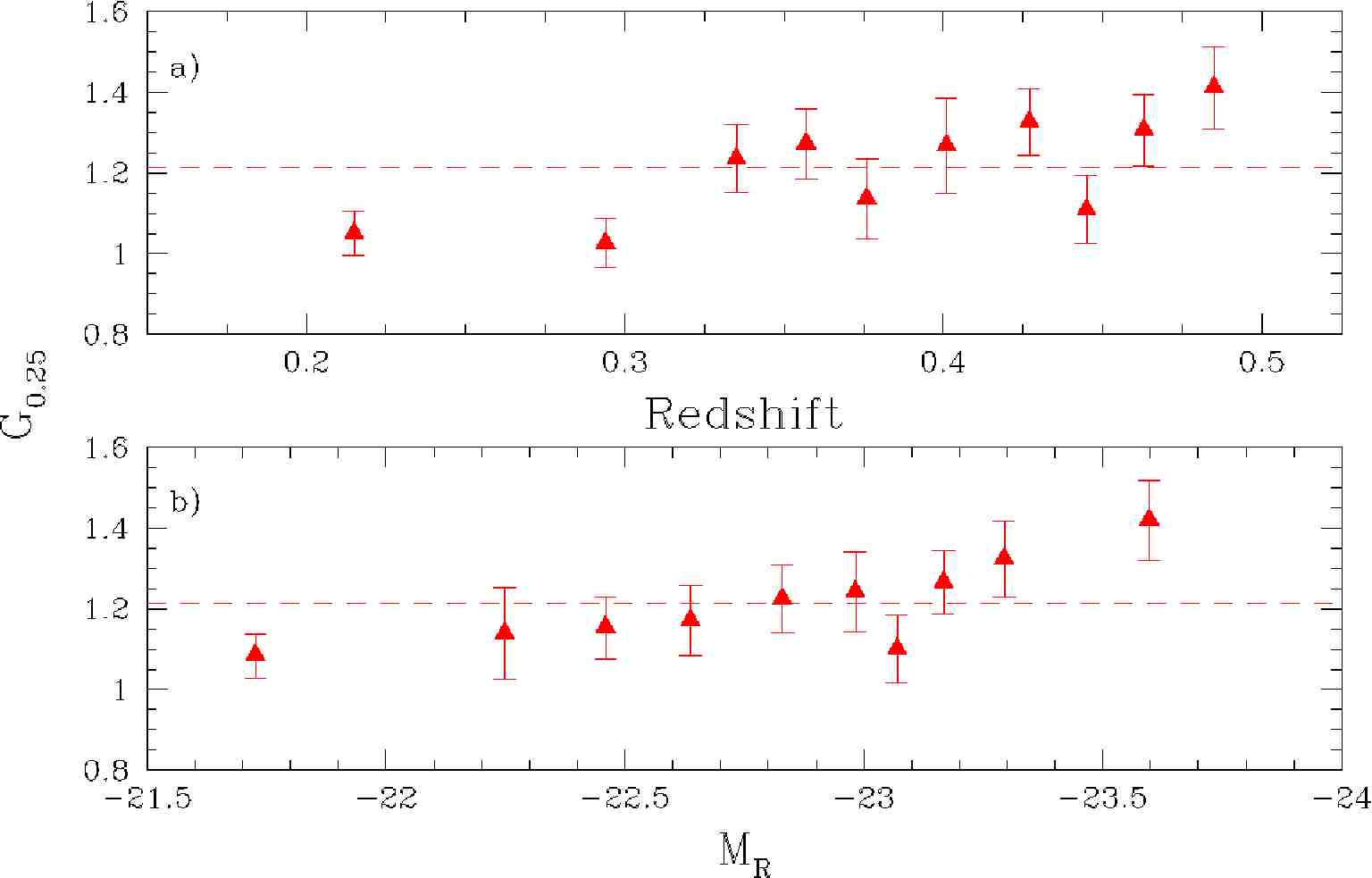} 
\caption{The dependence of the G$_{0.25}$ parameter on a) redshift, b) absolute magnitude of the galaxy for the matched control sample of inactive galaxies. The dashed red line shows the mean value of G$_{0.25}$ for the sample. In all cases the bins have been chosen to have the same number objects.}
\label{fig:g25depc}
\end{figure}

\begin{figure}
\centering
\includegraphics[width=\columnwidth]{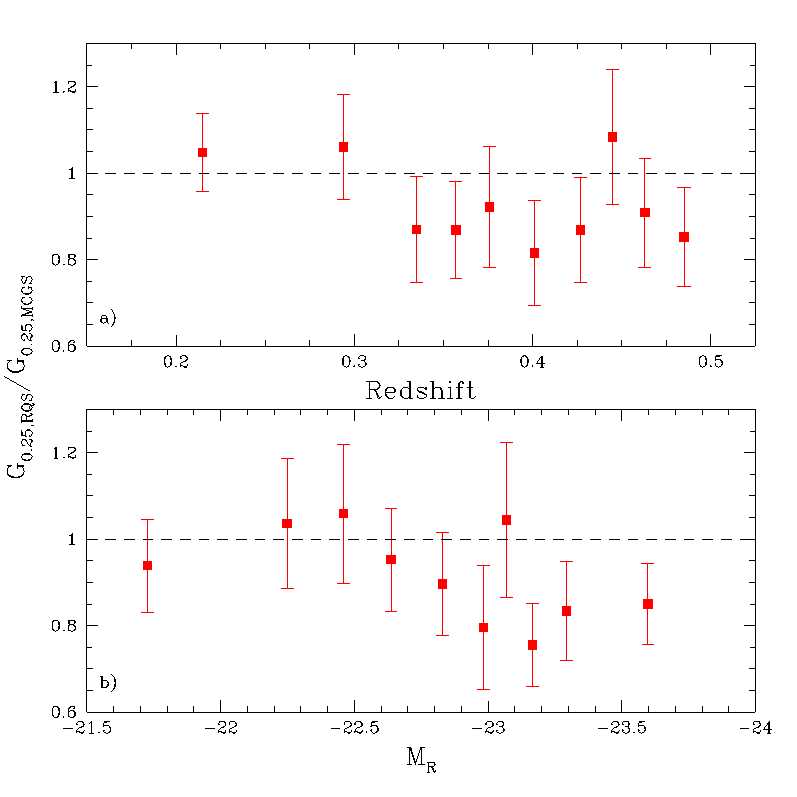}
\caption{The ratio G$_{0.25}$(RQS)/G$_{0.25}$(MCGS) as a function of a) redshift, b) absolute magnitude of the (host) galaxy. The dashed line shows the case where the overdensities on both samples are equal.}
\label{fig:g25ratio}
\end{figure}

Fig. \ref{fig:g25depc}a shows the G$_{0.25}$ parameter as a function of redshift for the MCGS. Unlike the RQS, the MCGS shows a trend of rising overdensity with increasing redshift. The best linear fit gives a slope of 1.09 $\pm$ 0.28, whereas the zero slope case is rejected with $P(\chi^2) < 0.05$. At redshifts below 0.3 the overdensity is marginal, having values consistent with no overdensity. This is slightly lower than the overdensity of $~1.11$ found for the RQS at the same redshifts, but the difference is not statistically significant. At $z > 0.3$ however, the overdensity has higher values and grows with redshift, reaching a value of 1.41 $\pm$ 0.10 for $z > 0.47$. The differences are better illustrated in Fig. \ref{fig:g25ratio}a, which shows the ratio G$_{0.25}$(RQS)/G$_{0.25}$(MCGS) at different redshifts. Even at the higher redshifts, the difference isn't significant.

\begin{table}
\caption{Linear weighted least-squares fit parameters for the matched quasar and control samples. The last column gives the $\chi^2$ probability of the fit.}
\begin{tabular}{|c|c|c|c|c|}
\hline
  \multicolumn{1}{|c|}{Sample} &
  \multicolumn{1}{c|}{Slope} &
  \multicolumn{1}{c|}{Intercept} &
  \multicolumn{1}{c|}{$\chi^2$} &
  \multicolumn{1}{c|}{P($\chi^2$)}\\
\hline
 RQS - z & $0.151 \pm 0.216$$^a$ & 1.059 & 6.13 & 0.63 \\
  & 0.0$^b$ & 1.110 & 6.62 & 0.47 \\
\hline
 MCGS - z & $1.090 \pm 0.280$$^a$ & 0.791 & 10.26 & 0.25 \\
  & 0.0$^b$ & 1.171 & 25.38 & $< 0.05$ \\
\hline
 RQS - $M_R$ & $-0.039 \pm 0.030$$^a$ & 0.232 & 11.77 & 0.16 \\
  & 0.0$^b$ & 1.113 & 13.53 & 0.06 \\
\hline
 RQS - $M_{R,host}$ & $-0.025 \pm 0.033$$^a$ & 0.560 & 13.46 & 0.10 \\
  & 0.0$^b$ & 1.126 & 14.04 & 0.05 \\
\hline
 MCGS - $M_R$ & $-0.135 \pm 0.044$$^a$ & -1.865 & 4.60 & 0.80 \\
  & 0.0$^b$ & 1.189 & 14.20 & $< 0.05$ \\
\hline
 RQS - $M_{BH}$ & $-0.018 \pm 0.039$$^a$ & 1.249 & 16.38 & $< 0.05$ \\
  & 0.0$^b$ & 1.095 & 16.59 & $< 0.05$ \\
\hline
\end{tabular}
\begin{list}{}
\item   $^{(a)}$ Fitted slope  \\
        $^{(b)}$ Zero-slope linear fit \\     
\end{list}
\label{tab:lsq}
\end{table}

\subsection{Dependence of environment on quasar and host galaxy luminosity}

To investigate the dependence of the environment on the luminosity of the quasar and of its host galaxy, the RQS was divided into 10 magnitude bins based on the absolute magnitude of the quasar as determined by \citet{falomo2014}, with each bin again containing the same number of objects. Fig. \ref{fig:g25depq}b shows the $G_{0.25}$ parameter for the quasars as a function of the absolute magnitude of the quasar, $M_{R,QSO}$, where the magnitudes are corrected for the galactic extinction and k-corrected to the R band rest frame. Again we did a weighed linear fit to the data to study the luminosity dependence, with the fit parameters shown in Table \ref{tab:lsq}. Our best linear fit, with a slope of -0.04 $\pm$ 0.03, and the zero slope fit are both found to be consistent with the data, although due to weak slope and large associated errors, we find no dependence of the overdensity on the luminosity of the quasar. 

Similarly in Fig. \ref{fig:g25depq}c we show the $G_{0.25}$ parameter for the RQS as a function of $M_{R,host}$, while Fig. \ref{fig:g25depc}b shows the same for the MCGS. For the quasar sample the best linear fit gives a slope of -0.03 $\pm$ 0.03, but the fit is not a particularly good one, with $P(\chi^2) = 0.10$. The zero slope fit to the data was similarly poor, with $P(\chi^2)$ value just above 0.05. The bad fits are due to the low luminosity objects with $M_{R,host} ~-21.5$; a better fit with a slope of 0.03 $\pm$ 0.04 and $P(\chi^2) = 0.35$ is obtained if the lowest luminosity objects are excluded, but even then no luminosity dependence is found. Hence we do not find significant dependence of the galaxy environment on the host galaxy luminosity, with the possible exception of the lowest luminosity objects ($M_{R,host} = -21.5$) that are not found in regions with significant overdensity. For the MCGS, on the other hand, the best linear fit with a slope of -0.14 $\pm$ 0.04 is found to be a good match to the data, with the case of no dependence rejected (having $P(\chi^2) < 0.05$). Fig. \ref{fig:g25ratio}b, shows the ratio G$_{0.25}$(RQS)/G$_{0.25}$(MCGS) as a function of $M_{R}$. This ratio is consistent with unity for all but the most luminous galaxies. A similar trend was also noted by \citep{serber2006}.

To test if the overdensity of MCGS depends on the redshift in our case (see Fig. \ref{fig:g25depc}a) we search for redshift luminosity correlation. This was done by selecting smaller subsamples from each redshift bins that all had similar luminosity distributions. Even with with a sample of quasars of similar luminosities at all redshifts, the dependence of the overdensity on redshift remained, with a best linear fit of slope 1.028 $\pm$ 0.312. 

However, the fact that the overdensity around non-active galaxies increases both with luminosity and redshift while no such dependence is found for the quasars could help to explain the slightly higher overall overdensity found for the control sample reported earlier in Section 4. This is because both the quasar and control samples have most of their targets at high redshifts and luminosities, where the overdensities for non-active galaxies are larger than those of quasars.

\subsection{Dependence of environment on black hole mass}

As galaxy formation and evolution are heavily influenced by their environments \citep[e.g.]{kormendy09} and the properties of the galaxies and their central black holes are likely linked \citep[e.g.]{ferrarese06}, we also investigate the dependence of the environment on the black hole mass. We adopted the BH mass measurements obtained by \citet{shen11}, who use FWHM of $H_{\beta}$ to estimate the virial BH mass for all quasars in SDSS DR7 \citep[see][for more details]{falomo2014}. Fig. \ref{fig:g25depq}d shows the G$_{0.25}$ parameter as a function of black hole mass, along with a best linear fit (solid black line). Again we found no significant dependence of the environment on BH mass, although the fit to these data is somewhat poor (see Table \ref{tab:lsq})

\subsection{Dependence of environment on radio luminosity}

\begin{figure}
\centering
\includegraphics[width=\columnwidth]{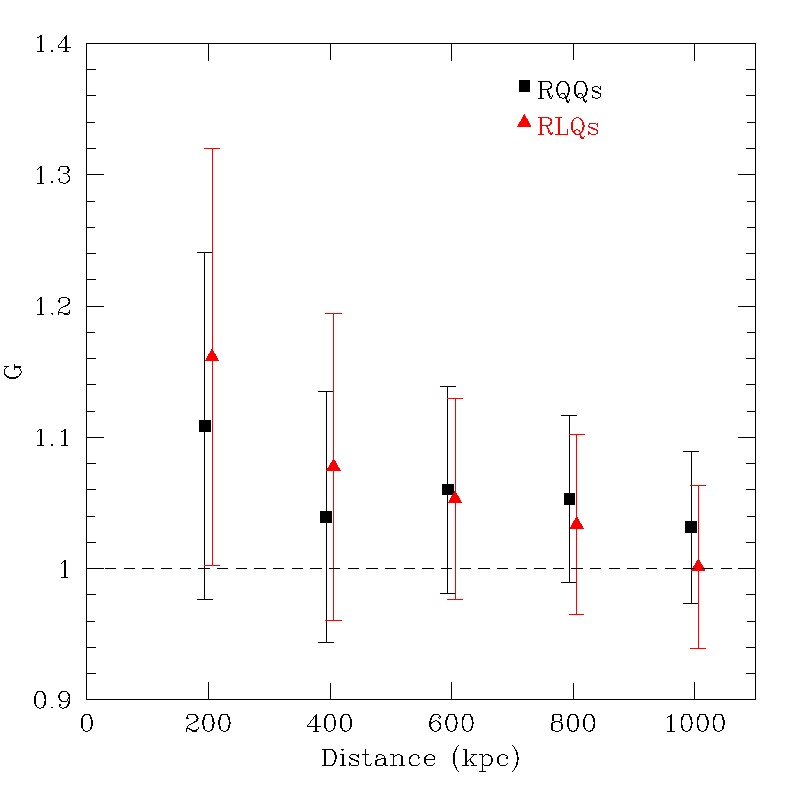}
\caption{Mean cumulative overdensity of the environments for the RQQ (black squares) and the RLQ (red triangles) subsamples as a function of distance. Points have been offset slightly horizontally for clarity.}
\label{fig:radio}
\end{figure}

Our quasar sample is dominated by radio quiet quasars and only 24 quasars in our sample are radio loud. Previous studies on the connection between the radio luminosity and the environmental galaxy density around quasars have shown mixed results, with studies like \citet{yee87} an \citet{ellingson91} finding the environments around RLQs to be denser than those around RQQs while studies by \citet{fisher96} and \citet{mclure01} find no difference between the environments of RLQs and RQQs. A more recent study by \citet{almeida13} found a dependence of radio luminosity of radio galaxies with the density of their environment. To check whether we find a dependence for the overdensity of the environment on the radio loudness. Since the number of RLQs was much smaller than that of RQQs, we built a subsample of 24 RQQs with a similar luminosity and redshift distribution as the RLQ subsample. Fig. \ref{fig:radio} shows the mean cumulative overdensity of the environments for the RLQ and RQQ subsamples as a function of distance from the quasar. No difference is found in the environments of RQQs and RLQs.

\section{Discussion}

The study presented here is the first to use the Stripe 82 co-added images to reach a depth of galaxy magnitudes of M*+2 up to the redshift of z = 0.5 and able to compare quasar environments to those of non-active galaxies selected by matching the galaxy properties to those of the quasar hosts. Previous studies of quasar environments have shown results both for and against the notion that quasars reside in environments denser than those of non-active galaxies. \citet{croom04}, for example, compared the environments of $~200$ quasars at $z < 0.3$ to those of normal galaxies at same redshifts using data from 2QZ, and found no difference between the two groups. On the other hand, studies such as \citet{serber2006} and \citet{strand2008} who used data from the DR3 and DR5 of the SDSS, respectively, report that quasar environments are denser than those of non-active galaxies. The control samples used in previous studies in general have not been selected to match the quasar samples in regards of the properties of the quasar hosts, instead being just restricted to have the same redshift range as the quasars \citep[e.g.][]{serber2006}. Others lack a control sample altogether \citep[e.g.][]{strand2008}, making it impossible to see how the presented results of quasar environments compare with those of regular galaxies. 

We focus our comparison to three studies, two of which use data from the SDSS. The study by \citet{serber2006} uses a large sample of quasars from the SDSS database at redshift range similar to the one used here, and also makes a comparison with a sample of regular galaxies. \citet{zhang13} use the SDSS Stripe 82 data to study quasars at redshifts larger than covered by this study ($0.6 < z < 1.2$). Additionally we briefly discuss a study by \citet{almeida13}, who studied the environments of radio galaxies and RQQs.

\subsection{Comparison with Serber et al. 2006}

\citet{serber2006} (S06 from now on) used data from SDSS DR3 to study the environments of 2028 quasars at $z < 0.4$, and comparing them to a control sample of $~10^5$ galaxies. However, unlike in our study, the quasar and galaxy samples were not matched with respect to luminosity. The density around the targets was determined by using the DR3 catalogs of photometric galaxies with magnitudes $14 \leq m_i \leq 21$, which means that their magnitude threshold is almost 2 magnitudes brighter than the one used in our study. In fact, as can be seen from Fig. \ref{fig:magthresh}, at z = 0.4 the galaxy catalogs used in S06 only contain relatively luminous galaxies up to an absolute magnitude of M*+1, missing a large number of fainter galaxies. 

In spite of this, the overall behaviour of the overdensity around quasars found in our study is consistent to that reported by S06. However, while we find the environments of inactive galaxies and quasars to have similar densities, S06 find the environments of quasars to be denser than those of L* galaxies, with the overdensity being largest at the smallest separations. For comparison, we have presented the results from Table 1 of S06 in the last two columns of our Table \ref{tab:ovden}. The value of the overdensities themselves are different from our results; we find an overdensity of $1.12\pm0.04$ within 250 kpc for the RQS, while S06 reports a value of $1.57\pm0.03$.

For the control sample, the results from S06 are not directly comparable to ours, as they only cite the overdensity around L* galaxies, whereas our control sample value is an average for the whole sample. For comparison, we calculated the overdensity for a subsection of 53 galaxies from the MCGS around the absolute magnitude adopted in S06 for L* galaxies. We find an overdensity of $1.15\pm0.07$ at a distance of 250 kpc, which is lower than the value of $1.235\pm0.004$ reported by S06, and agrees very well with the quasar overdensity. Thus, we find our quasar environments to correspond to those of L* galaxies while S06 finds quasar environments to have similar densities to 2L* galaxies.

The differences in the overdensities could be partly explained by the fainter magnitude limit used in this study; by including fainter background galaxies, both environmental and background galaxy density grow by the same amount, reducing the ratio of the two densities. To study how much this affects our results, we set our magnitude threshold to $m_i = 21$, as used by S06, and repeated the analysis. We find an overdensities of $1.17\pm0.09$ and $1.29\pm0.10$ for the RQS and MCGS within a distance of 250 kpc. For L* galaxies we find an overdensity of $1.15\pm0.13$, still clearly lower than that reported by S06. At smaller separations ($<$ 100 kpc) we find overdensities of $1.17\pm0.20$ and $1.46\pm0.24$ for the RQS and MCGS, respectively, while for the L* galaxies we get an overdensity of $1.62\pm0.34$, which agrees with what S06 found within the uncertainties. We notice that due to the large uncertainties, the overdensity around L* galaxies at distances $<$ 100 kpc is also consistent with the overdensity around quasars. Therefore while the brighter magnitude limit used by S06 could account for the difference in overdensity for the L* galaxies, it can not explain the larger differences found in the quasar environments.

No dependence of the overdensity on redshift or the luminosity of the quasar was found in our study. S06 also find no redshift dependence (see Fig. 1 of their paper), but they did find that quasars brighter than $M_i = -23.3$ have denser environments than their fainter counterparts. However this effect was found to occur at the scales of $< 100 kpc$ and disappears at distances of larger than 100kpc. At distances smaller than 100kpc we have very few galaxies, and the variations in the environmental density between quasars are large. This combined with our smaller quasar sample lead to poor statistics, and large errors at small separations.

For our inactive galaxies, we find the luminosity-overdensity relation to be less steep than that found by S06, with galaxies one magnitude brighter than L* having an overdensity of $1.74\pm0.18$, compared with $1.36\pm0.13$ for L* galaxies at a distance of 100kpc. At a distance of 250kpc, we find an enhancement of overdensity by a factor of 1.17 for the same luminosity range. The relationship found between the overdensity and the luminosity of the galaxies is much weaker than that found by S06. Overall, we find smaller overdensities around both quasars and inactive galaxies than those reported by S06, and we also find no difference between the two classes of objects contrary to S06. The differences in the overdensities can partly be explained by the deeper magnitude limit used in our study, especially for the L* galaxies, but the cause of the large differences in the quasar environments remains unknown.

\subsection{Comparison with Zhang et al. 2013}

\citet{zhang13} (called Z13 from now on) used data from the SDSS Stripe 82 to study the environments of 2300 quasars at redshifts $0.6 < z < 1.2$ on scales of $0.05 < r_p < 20 h^{-1} Mpc$. As the redshift range differs from that used in our study, no direct comparison between the two can be done. However, it is in principle possible to use Z13 to compare how the density changes at redshifts higher than those in our study, while keeping in mind that the absolute magnitudes of the quasars studied by Z13 are on average 0.5-1.5 magnitudes brighter than ours. Z13 uses the SDSS DR7 Stripe 82 calibrated object catalog to study the quasar environments, but they don't exclude stars in the catalog from their sample of "galaxies". This combined with the objects in the Stripe 82 catalogs at magnitudes fainter than $m_i = 23$ that we detected and classified as "noise" leads to "background" densities that are intrinsically 10-50\% higher, than those found in our study.

\begin{figure*}
\centering
\includegraphics[width=\textwidth]{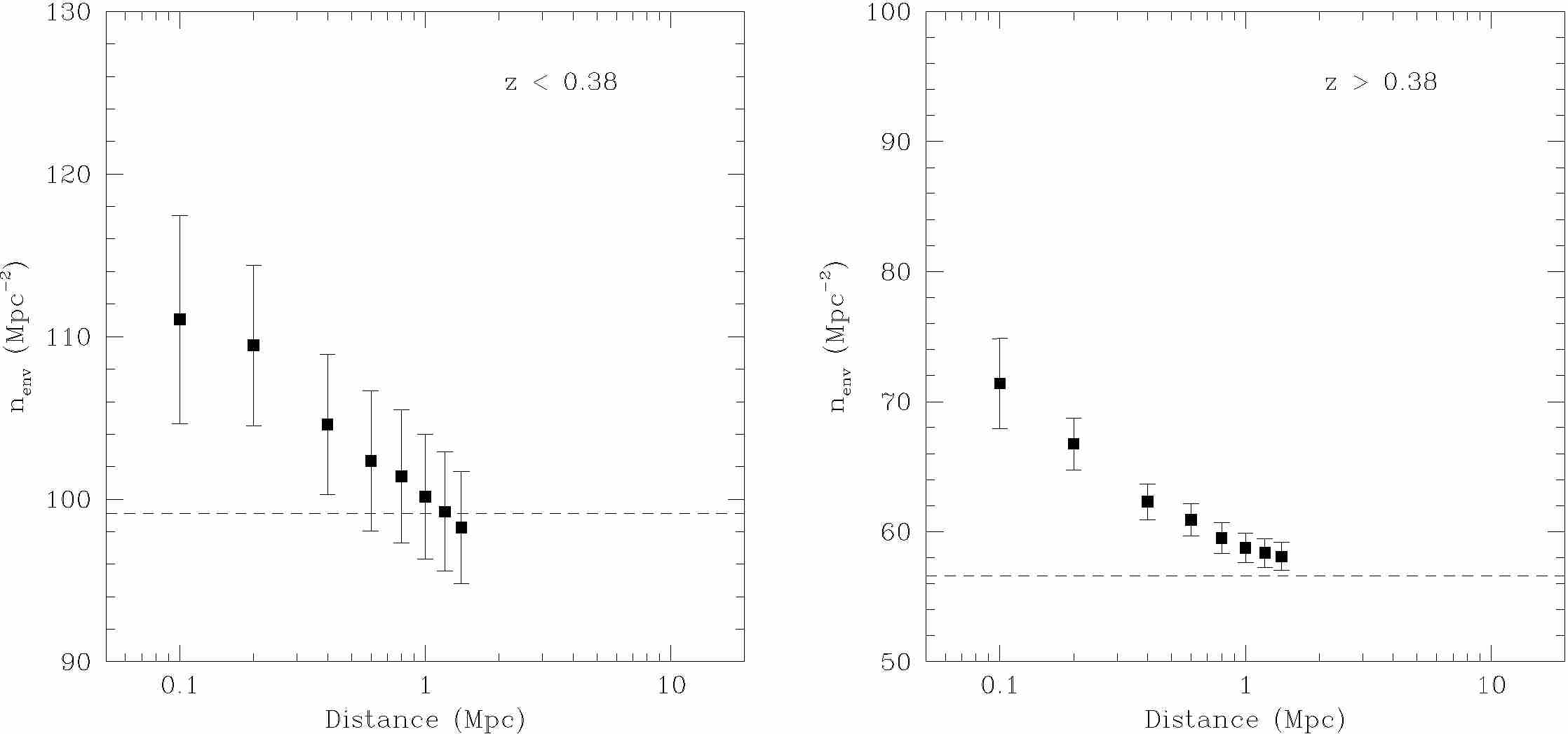}
\caption{The averaged galaxy number density as a function of projected distance from the quasar for quasars at a) $z < 0.38$ and b) $z > 0.38$. The redshift cut was made to ensure each subsample had roughly the same number of objects. The dashed black line shows the average background galaxy density.}
\label{fig:zhangcom}
\end{figure*}

Z13 measured the galaxy overdensity around the quasars using the clustering amplitude, $r_0$ as opposed to the overdensity of the environment itself. However, Fig. 6 of Z13 shows a plot of averaged galaxy number density as a function of projected distance from the quasar for the different redshift intervals. Fig. \ref{fig:zhangcom} shows a similar plot for our data, split into "low-" and "high-redshift" groups, although our data covers a smaller range in projected distance. 

We converted the Z13 galaxy background densities (given in $Mpc^{-2}$) in each of the redshift bins to that of a constant angular area. As the average background counts of galaxies is taken for objects of different redshifts, we used the average redshift of the bin to calculate the new densities. For our data, we find an average background density of ~5.9 $arcmin^{-2}$ for both redshift groups. For the Z13 data we get values of 9.9, 7.9 and 6.9 $arcmin^{-2}$ for the low-, mid- and high-redshift subsamples. The higher values of the background densities can be expected due to the inclusion on stars and background noise in the catalogs, but the significant variation of the background density between different redshift samples remains unexplained.

We find no dependence of galaxy density on redshift, while Z13 claim to find an increase in the average clustering amplitude with redshift. However, the values given in Table 2 of Z13 show that the clustering amplitude of all three redshift bins agree within their uncertainties, making the redshift dependence very marginal.

We do not find a dependence of the galaxy density on the luminosity of the quasar at redshift $z < 0.5$, and Z13 reports similar results at their redshift range. This implies that the lack of luminosity dependence found by our study extends to quasars of higher luminosities, at least at redshifts of $0.6 < z < 1.2$. Finally, contrary to our results, Z13 finds quasars with heavier central black holes to have larger clustering amplitudes than the ones with lighter black holes. We note again that the values of the clustering amplitude for the high- and low black hole mass subsamples given in Table 3 of Z13 are consistent with each other within their uncertainties.

\subsection{Comparison with Ramos Almeida et al. 2013}

\citet{almeida13} (RA13) studied the environments of 46 radio galaxies at redshifts $0.05 < z < 0.7$, 20 RQQs at $0.3 < z < 0.41$ and 107 non-active early-type galaxies at $0.2 < z < 0.7$. Due to the small redshift range of the RQQ sample, a smaller subsample of 19 radio galaxies with $0.2 < z < 0.7$ was used for comparison. Similarly, a subsample of 51 non-active galaxies with $0.3 < z < 0.41$ derived from the full galaxy sample was used for comparing with the RQQs. RA13 used SExtractor to generate their own galaxy catalogs, by including all objects detected by the program with a classification $< 0.85$ as opposed to the more conservative limits used in this study.

The density of the environments was studied by using the spatial clustering amplitude, $B_{gq}$ \citep{longair79}. The clustering amplitude was calculated for a distance of 170kpc from the target for all three samples. We calculated the $B_{gq}$ parameters for the RQS and MCGS, and found values of $B_{gq,RQS} = 34\pm13$ and $B_{gq,MCGS} = 56\pm15$ for the two samples. RA13 found that the environments of the RQQs are not significantly different from those of non-active galaxies, with average clustering amplitudes of $B_{gq} = 151\pm76$ and $B_{gq} = 79\pm26$ for the RQQs and non-active galaxies, respectively.

For our RLQ and RQQ samples, we find $B_{gq,RLQ} = 66\pm47$ and $B_{gq,RQQ} = 27\pm35$, which are consistent with each other. However, RA13 find an average clustering amplitude of $B_{gq} = 395\pm84$ for the radio galaxies, which is much higher than that found for RQQs. Although the significance of the difference is only $2\sigma$, this strongly implies that radio galaxies reside in denser environments than radio quiet. We note that the radio luminosities of our RLQs are much lower than those of the radio galaxies studied in RA13, which might explain why we detect no density enhancement in comparison to our RQQs.

\section{Conclusions}

We have used SDSS Stripe 82 data to study the $<1 Mpc$ environments of low redshift ($0.1<z<0.5$) quasars, and compared them with the environments of galaxies in the control sample of non active galaxies. The environments were studied by measuring the number density of galaxies within a projected distance of 200 kpc to 1 Mpc from the quasars, and then dividing this by the number density of galaxies in the background to calculate the overdensity of the region. The overdensities associated with the quasar environments were then compared with those of a sample of control galaxies well matched in both the redshift and the galaxy luminosity.

We find the following results:
\begin{enumerate}
\item Quasars are on average found associated with small group of galaxies. The overdensities of galaxies are mainly observed in the closest ($<$ 200 kpc) region around the source and vanish at distance of 1 Mpc.

\item No statistically significant difference is found between the overdensities around the quasars and the inactive galaxies at any separation.

\item No dependence of the overdensity on redshift, quasar luminosity, the luminosity of the host galaxy, black hole mass or the radio luminosity was found. The result is the same for both our full and the matched samples.
\end{enumerate}

We compared our results to those of previous studies, in particular the ones by \citet{serber2006} who studied low redshift quasar environments ($z < 0.4$) and \citet{zhang13} who focused on redshift of $0.6 < z < 1.2$. We find that the galaxy overdensity is independent of either redshift or luminosity. For our sample of inactive galaxies we find a trend of higher overdensities with increasing luminosity. Contrary to \citet{serber2006}, we find the environments of quasars are similar to those of inactive galaxies. Additionally, the overdensities we find are lower for both the quasars and galaxies than those of \citet{serber2006}.

The fact that we find no significant difference between the environments of quasars and non-active galaxies suggests that the link between the quasar activity and the environment of the quasars is less important than believed for fueling and trigerring the activity. This also points to smaller importance of (major) mergers than expected in triggering the quasar activity, and that secular evolution (e.g. disk instabilities) may play an important role. Similarity between the environments of quasars and non-active galaxies could thus indicate that the quasar phase is a common event in the life cycle of a massive galaxy and does not depend significantly on the local environment at least on the scales studied here. We also find no dependence of the radio loudness of the quasar on the overdensity, in agreement with studies by \citet{fisher96} and \citet{mclure01}. However, as our sample of RLQs is very small, and their radio luminosities quite low, we cannot make rule out the possibility connection between the environment and radio activity, such as the one found in \citet{almeida13}.

A detailed study of morphology, peculiarities of quasar hosts and colours of this large sample of AGN could provide further clues to understanding the link between nuclear activity and the processes fueling and triggering it. This will be explored in a future paper of this series (paper III, Bettoni et al. in prep). Finally, the colours of the galaxies in the environments of the quasars in this sample will be explored in paper IV of the series (Karhunen et al. in prep).

\section{Acknowledgments}

K. K. acknowledges financial support from the Finnish Academy of Science and Letters (Vilho, Yrj\"o and Kalle V\"ais\"al\"a Foundation).

Funding for the SDSS and SDSS-II has been provided by the Alfred P. Sloan Foundation, the Participating Institutions, the National Science Foundation, the U.S. Department of Energy, the National Aeronautics and Space Administration, the Japanese Monbukagakusho, the Max Planck Society, and the Higher Education Funding Council for England. The SDSS Web Site is http://www.sdss.org/.

The SDSS is managed by the Astrophysical Research Consortium for the Participating Institutions. The Participating Institutions are the American Museum of Natural History, Astrophysical Institute Potsdam, University of Basel, University of Cambridge, Case Western Reserve University, University of Chicago, Drexel University, Fermilab, the Institute for Advanced Study, the Japan Participation Group, Johns Hopkins University, the Joint Institute for Nuclear Astrophysics, the Kavli Institute for Particle Astrophysics and Cosmology, the Korean Scientist Group, the Chinese Academy of Sciences (LAMOST), Los Alamos National Laboratory, the Max-Planck-Institute for Astronomy (MPIA), the Max-Planck-Institute for Astrophysics (MPA), New Mexico State University, Ohio State University, University of Pittsburgh, University of Portsmouth, Princeton University, the United States Naval Observatory, and the University of Washington.

\section{Appendix A - Comparison between the SDSS catalog and SExtractor catalog}

\begin{figure}
\centering
\includegraphics[width=\columnwidth]{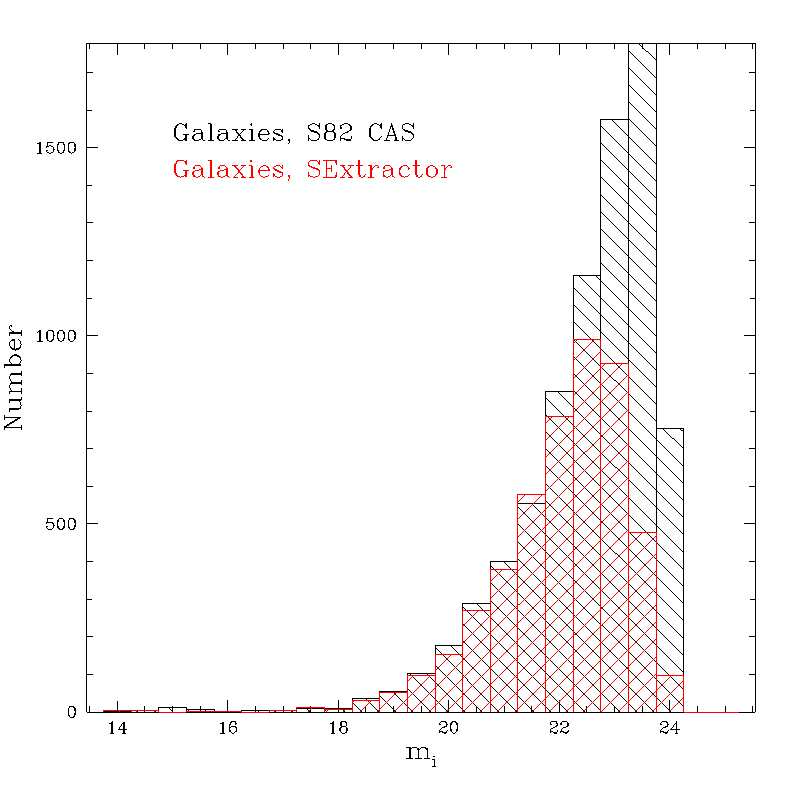}
\caption{Magnitude distribution of galaxies in the field of SDSS J013023.51+000551.7 from the Stripe 82 catalogs (black) and the SExtractor catalogs (red). }
\label{fig:obj250com}
\end{figure}

We compared the available SDSS Stripe 82 galaxy database with catalogs generated by SExtractor by analysing the co-added images. The comparison was done on three separate fields, each associated with one of the quasars in our sample (SDSS J203657.28+000144.3, SDSS J013023.51+000551.7 and SDSS J023922.87-000119.5). The Stripe 82 catalogs were generated using the Stripe 82 Catalog Archive Server\footnote{\url{http://cas.sdss.org/stripe82/en/}} by including all the objects classified as galaxies within the studied field with apparent magnitudes of $m_i < 24$. The Stripe 82 catalogs and the SExtractor catalogs were then matched with respect to object position to determine how the classifications of the objects might vary between the two catalogs. Fig. \ref{fig:obj250com} shows the apparent magnitude distribution of galaxies for the two catalogs, with the S82 galaxies in black and SExtractor galaxies in red. The two catalogs are well matched for objects with $m_i < 22.5$, where more than 90\% of the galaxies are found in both catalogs. At fainter magnitudes, however, the amount of galaxies detected by SExtractor drops quickly. Including objects classified as "unknown" by SExtractor, improves the situation slightly, but still at least 50\% of the objects in the S82 catalog with $m_i > 23$ are completely missed.

\begin{figure}
\centering
\includegraphics[width=\columnwidth]{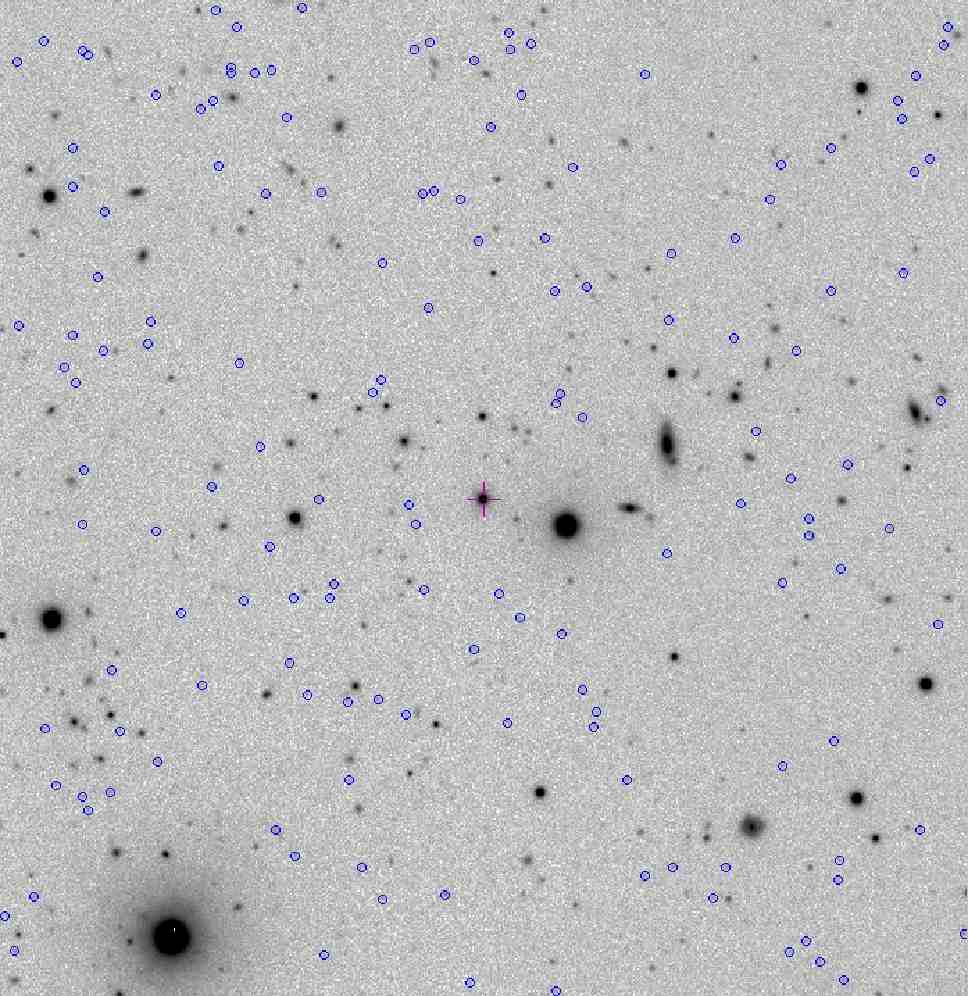}
\caption{A picture showing the galaxies with $m_i > 23$ (blue circles) in the Stripe 82 catalogs in the field of SDSS J013023.51+000551.7 (position marked by the purple crosshairs).}
\label{fig:qso_23}
\end{figure}

\begin{figure}
\centering
\includegraphics[width=\columnwidth]{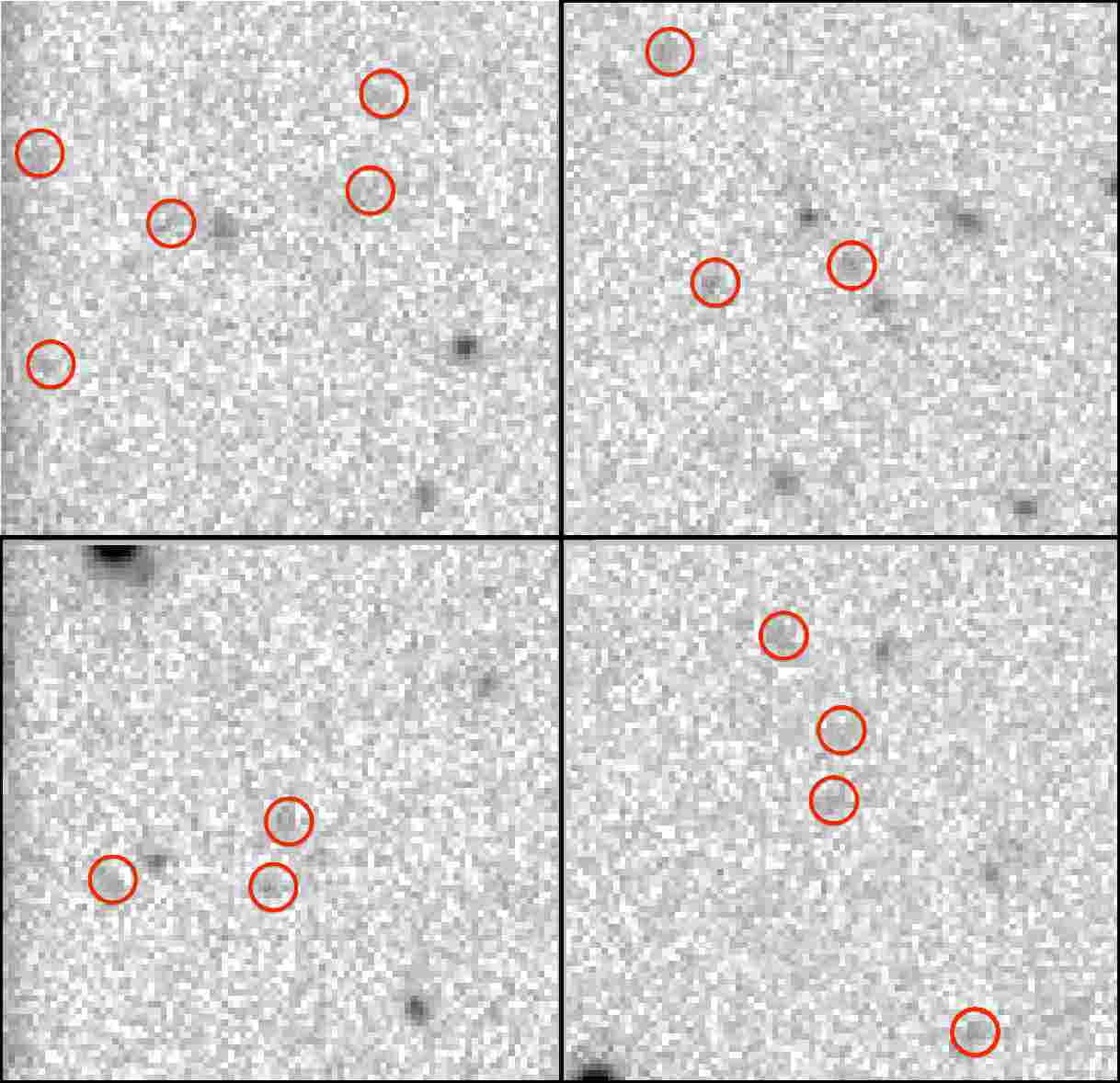}
\caption{Close-ups of a number of galaxies with $m_i > 23$ (red circles) in the Stripe 82 catalogs in the field of SDSS J013023.51+000551.7.}
\label{fig:4panels}
\end{figure}

To study the reason for the "missing" faint objects, we performed a visual inspection of the frames. The visual inspection of faint objects ($m_i > 23$) shows that they are mainly background noise which was either undetected by SExtractor or classified as an "unknown" object. This is illustrated in Fig. \ref{fig:qso_23}, which shows the field containing quasar SDSS J013023.51+000551.7, with all galaxies in the S82 catalog with $m_i > 23$ marked with blue circles. Fig. \ref{fig:4panels} shows four close-up views of the field, showing the faint S82 catalog objects in more detail.

\newpage
\null

\end{document}